\documentclass[sigconf]{acmart}

\definecolor{darkgreen}{rgb}{0.0, 0.56, 0.0}
\definecolor{amethyst}{rgb}{0.6, 0.4, 0.8}
\definecolor{blue-violet}{rgb}{0.54, 0.17, 0.89}

\usepackage{tikz}
\usepackage{pgf-pie}
\usepackage{pgfplots}
\usepackage{pgfplotstable}
\usetikzlibrary{patterns}
\usepgfplotslibrary{colormaps}
\usetikzlibrary{pgfplots.colormaps}
\usepackage{xcolor,colortbl}
\usepackage{multirow}
\usepackage{makecell}
\usepackage{longtable}
\usepackage{graphicx}

\pgfplotsset{
    draw group line/.style n args={5}{
        after end axis/.append code={
            \setcounter{groupcount}{0}
            \pgfplotstableforeachcolumnelement{#1}\of\datatable\as\cell{%
                \def\temp{#2}
                \ifx\temp\cell
                    \ifnum\thegroupcount=0
                        \stepcounter{groupcount}
                        \pgfplotstablegetelem{\pgfplotstablerow}{[index]0}\of\datatable
                        \coordinate [yshift=#4] (startgroup) at (axis cs:\pgfplotsretval,0);
                    \else
                        \pgfplotstablegetelem{\pgfplotstablerow}{[index]0}\of\datatable
                        \coordinate [yshift=#4] (endgroup) at (axis cs:\pgfplotsretval,0);
                    \fi
                \else
                    \ifnum\thegroupcount=1
                        \setcounter{groupcount}{0}
                        \draw [
                            shorten >=-#5,
                            shorten <=-#5
                        ] (startgroup) -- node [anchor=north] {#3} (endgroup);
                    \fi
                \fi
            }
            \ifnum\thegroupcount=1
                        \setcounter{groupcount}{0}
                        \draw [
                            shorten >=-#5,
                            shorten <=-#5
                        ] (startgroup) -- node [anchor=north] {#3} (endgroup);
            \fi
        }
    }
}
\makeatother

\pgfplotstableread{
1	29	0	18	9	0	3   1   1   2
2 	9	9	5	8	11	8   8   1   2
3	7	7	13	7	8	9   2   0   4
4	2	7	16	7	6	12  1   0   0
5	6 	1  	9	7	1	23  0   0   0
}\datatable

\pgfplotstableread{
1   17	22  0   15  9
2   6	17  10  10  13
3   3	15  19  5   11
4   0   6   20	3   6
5   0	0   1   0   0
}\loadedtable

\AtBeginDocument{%
  \providecommand\BibTeX{{%
    \normalfont B\kern-0.5em{\scshape i\kern-0.25em b}\kern-0.8em\TeX}}}
\copyrightyear{2023}
\acmYear{2023}
\setcopyright{rightsretained}

\acmConference[AIES '23]{AAAI/ACM Conference on AI, Ethics, and Society}{August 8--10, 2023}{Montréal, QC, Canada}
\acmBooktitle{AAAI/ACM Conference on AI, Ethics, and Society (AIES '23), August 8--10, 2023, Montréal, QC, Canada}
\acmDOI{10.1145/3600211.3604666} \acmISBN{979-8-4007-0231-0/23/08}

\usepackage{etoolbox}
\makeatletter
\patchcmd{\maketitle}{\@copyrightpermission}{
   \begin{minipage}{0.3\columnwidth}
     \href{https://creativecommons.org/licenses/by-nc-sa/4.0/}{\includegraphics[width=0.90\textwidth]{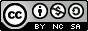}}
   \end{minipage}\hfill
   \begin{minipage}{0.7\columnwidth}
     \href{https://creativecommons.org/licenses/by-nc-sa/4.0/}{This work is licensed under a Creative Commons Attribution-NonCommercial-ShareAlike International 4.0 License.}
   \end{minipage}
}{}{}

\begin{document}

\title{Evaluating Biased Attitude Associations\\of Language Models in an Intersectional Context}

\author{Shiva Omrani Sabbaghi}
\email{somrani@gwu.edu}
\affiliation{%
  \institution{George Washington University}
  \city{Washington, District of Columbia}
  \country{USA}}

\author{Robert Wolfe}
\email{rwolfe3@uw.edu}
\affiliation{%
  \institution{University of Washington }
  \city{Seattle, Washington}
  \country{USA}}

\author{Aylin Caliskan}
\email{aylin@uw.edu}
\affiliation{%
  \institution{University of Washington }
  \city{Seattle, Washington}
  \country{USA}}

\begin{abstract}

Language models are trained on large-scale corpora that embed implicit biases documented in psychology. Valence associations (pleasantness/unpleasantness) of social groups determine the biased attitudes towards groups and concepts in social cognition. Building on this established literature, we quantify how social groups are valenced in English language models using a sentence template that provides an intersectional context. We study biases related to age, education, gender, height, intelligence, literacy, race, religion, sex, sexual orientation, social class, and weight. We present a concept projection approach to capture the valence subspace through contextualized word embeddings of language models. Adapting the projection-based approach to embedding association tests that quantify bias, we find that language models exhibit the most biased attitudes against gender identity,  social class, and sexual orientation signals in language. We find that the largest and better-performing model that we study is also more biased as it effectively captures bias embedded in sociocultural data. We validate the bias evaluation method by overperforming on an intrinsic valence evaluation task. The  approach enables us to measure complex intersectional biases as they are known to manifest in the outputs and applications of language models that perpetuate historical biases. Moreover, our approach contributes to design justice as it studies the associations of groups underrepresented in language such as transgender and homosexual individuals.
\end{abstract}



\maketitle

\section{Introduction}
\label{sec:intro}
Static word embeddings \cite{pennington-etal-2014-glove, bojanowski2017enriching} are known to reflect the semantics and biases of the populations that produce the data on which they are trained \cite{Caliskan_2017,toney2020valnorm,caliskan2020social}. While problematic for their use in machine learning applications which are affected by these biased features \cite{zhao-etal-2017-men, leino2019featurewise}, static word embeddings have also allowed for the development of new social scientific approaches to studying societal norms and biases \cite{GargE3635, LewisMolly2020Gsar, hamilton-etal-2016-diachronic, doi:10.1177/0003122419877135}. However, static word embeddings have been replaced as the dominant representational paradigm in natural language processing (NLP) by language models \cite{peters2018deep,devlin-etal-2019-bert, Radford2018ImprovingLU, radford2019language,NEURIPS2020_1457c0d6}, which form contextualized word embeddings, dynamic representations of words that undergo change over the course of the neural network based on the words which occur around them. Prior work suggests that, as this process of "contextualization" occurs, a contextualized representation becomes more semantically similar to the words which occur in context around it \cite{wolfe2022vast}.

How can a principled and generalizable test for social bias, including intersectional bias, be designed for such dynamic representations? The present research proposes that, rather than studying changes in the representation of a certain word being evaluated for bias, one might instead look to the effects that a biased word has on its surrounding context. That is, instead of finding ways to compensate for the effects of contextualization when assessing bias, one can use the dynamic properties of language models to design a generalizable bias assessment method specifically suited to the paradigm of contextualization.

The first challenge in designing a bias test for contextualized word embeddings, however, is that they are not easy to analyze using common mathematical methods for measuring similarity between word embeddings, such as cosine similarity. While prior work has used principal component analysis (PCA) of subtracted vectors to find the dimension that maximizes the variance between biased representations \cite{10.5555/3157382.3157584}, contextualized word embeddings are known to contain high-magnitude neurons which are often not semantic in nature \cite{wolfe2022vast,timkey2021all}, preventing the development of a generalizable method for assessing semantic biases based on PCA.

The present research addresses this problem by using a maximum margin support vector classifier to learn a semantic property of the contextualized word embedding space: namely, the valence (pleasantness vs. unpleasantness) subspace \cite{osgood1957measurement}, onto which embeddings can be projected to measure their semantic properties. In social psychology, valence associations determine the biased attitudes towards social groups \cite{greenwald1998measuring}. For example, are European American men or African American women perceived more positively valenced? Our method for isolating semantics in contextual spaces also allows for the introduction of a generalizable statistical test to quantify bias in language models by measuring the effects of contextualization. This work applies these methods to five language models (GPT-Neo, XLNet, ALBERT, RoBERTa, and T5) of varying architectures and demonstrates the ability to measure both contextualized word embedding semantics and bias in language models.

Code and data are made public at \url{https://github.com/shivaomrani/LLM-Bias}. The contributions of this research are outlined below:

\begin{enumerate}
    \item A method based on learning a maximum margin subspace to learn the valence subspace of an embedding space is introduced for isolating semantics in the highly contextual and anisotropic upper layers of contextualizing language models. Across five evaluated language models and without resort to pooling methods or postprocessing the contextualized embedding space, the approach is demonstrated to be robust to the geometry of contextualized embedding spaces, and outperforms a cosine similarity based method in the upper layers of every model. In GPT-Neo \cite{gao2020pile}, scores on the ValNorm intrinsic evaluation task \cite{toney2020valnorm}, which measures the correlation (Pearson's $\rho$) of human-rated valence with valence associations in models, fall to $0.56$ in the top layer of the model when using cosine similarity; with the maximum margin method, the score remains high, at $0.81$. A similar result is obtained for the four other language models studied, indicating the utility of the method for studying semantics in highly contextual and anisotropic embedding spaces.

    \item A statistical bias measurement based on the Word Embedding Association Test (WEAT) \cite{Caliskan_2017} is introduced to study differential biases arising from the process of contextualization in language models. The word "person" is placed into generated intersectional contexts with a wide variety of words reflecting social groups. "Person" is contextualized by these contexts, and its embedded representation is obtained from the top layer of a language model. The differential bias between two words is obtained by measuring their effect on the contextualized representation of the word "person" when placed in otherwise identical contexts, as measured based on the projection product with valence (pleasantness vs. unpleasantness) subspace. The method captures a wide variety of biases in language models related to age, education, gender, height, intelligence, literacy, race, religion, sex, sexual orientation, social class, and weight. The results reveal pronounced biases across five language models associated with gender identity  (average effect size $d = 0.60$ - "cisgender" and "transgender"), social class (average effect size $d = 0.48$ - "affluent" and "destitute"), and sexual orientation (average effect size $d = 0.42$ - "heterosexual" and "homosexual").
    \item A method is introduced for studying biases without need for a binary, differential test. A permutation is used to generate a large sample of sentences that include social group signals in an intersectional context, each ending with the word "person." The embedded representation of person is computed from each sentence, and the projection product is obtained with the maximum margin subspace. The top 10\% most pleasant sentences are returned, and the top 10\% most unpleasant sentences are returned. In GPT-Neo, more than 90\% of the most pleasant sentences contain the word "heterosexual," while more than 99\% of the most unpleasant phrases contain the word "homosexual," again reflecting significant biases related to sexual orientation. Similar biases exist for gender identity in GPT-Neo, with more than 70\% of the most pleasant phrases including the word "cisgender," and more than 93\% of the most unpleasant phrases including the word "transgender."
\end{enumerate}

The results of this research have implications both for the study of bias in AI, where researchers might employ the bias evaluation method to analyze language models for a wide range of intersectional biases by learning subspaces separating conceptual categories or evaluate the effectiveness of bias mitigation approaches, and for the social sciences, which might employ this approach to study the human biases encoded into machines.

\section{Related Work}
\label{sec:relwork}

The present research contributes new methods for measuring semantic norms and bias in contextualized word embeddings. This section reviews related work on the measurement of semantics and bias in static and contextualized word embeddings.

\subsection{Static and Contextualized Word Embeddings}

Word embeddings are dense, continuous-valued vector representations of words used to encode a statistical model of human language \cite{bengio2003neural}. Static word embeddings such as those formed using the GloVe \cite{pennington-etal-2014-glove} and fastText \cite{mikolov2018advances} algorithms are trained on the co-occurrence statistics of words in a language corpus, and encode the semantic properties of language \cite{mikolov2013distributed}, such that algebraic operations on embedded representations can be used to solve analogical tasks \cite{mikolov-etal-2013-linguistic}. Static word embeddings are known to encode societal attitudes and implicit and explicit biases of the population which produced the linguistic data on which they are trained \cite{manzini-etal-2019-black, GargE3635, swinger2019biases}. While identifying and mitigating bias in word embeddings is a noteworthy area of study due to the propagation of these biases in downstream natural language processing (NLP) applications \cite{zhao-etal-2017-men, leino2019featurewise, caliskan2021detecting, cheng2023marked, mei2023bias}, the encoding of population-level human attitudes in word embeddings also allows them to be used as a statistical tool for studying bias, languages, societies, and historical events \cite{omrani2022measuring, GargE3635, LewisMolly2020Gsar, toney2021automatically, wolfe2022detecting, hamilton-etal-2016-diachronic, doi:10.1177/0003122419877135, charlesworth2022historical}. 

Despite their widespread usefulness for both computer science and the social sciences, static word embeddings have a central limitation, in that they collapse all of the senses of a word into a single vector representation. Contextualizing language models such as ELMo \cite{peters-etal-2018-deep}, BERT \cite{devlin-etal-2019-bert}, and the GPT family of models \cite{Radford2018ImprovingLU,radford2019language,NEURIPS2020_1457c0d6,gao2020pile} overcome this limitation by forming contextualized word embeddings, which incorporate information from surrounding words, such that the representation of a word depends on the context in which it appears. Therefore, while polysemes and homographs (words with the same spelling but different meaning) share the same representation in static word embeddings, contextualized word embeddings capture semantic differences based on context and alter a word's representation to reflect the sense in which it is used \cite{soler2021let}. However, polysemes and homographs are not the only words which change representation as they are processed in a contextualizing language model. \citet{ethayarajh-2019-contextual} shows that stopwords and articles are some of the most context-sensitive words in models like GPT-2, while \citet{wolfe2022vast} demonstrate that contextualized word embeddings from seven language models become more semantically similar to the words that occur around them as they are processed in the model.

This suggests that a test of social attitudes and biases encoded in language models might be designed based on the effect a word has on the embedded representations of the words which occur around it. However, contextualized word embeddings have their own limitation: anisotropy, or directional uniformity \cite{ethayarajh-2019-contextual}. Because language models are trained on a wide variety of objectives such as next-word prediction \cite{Radford2018ImprovingLU} and masked-word prediction \cite{devlin-etal-2019-bert}, the geometric structure of contextualized word embeddings may reflect properties useful to performing a model's pretraining task, but detrimental for assessing embedding semantics using methods such as cosine similarity \cite{wolfe2022vast}. Recent research proposes methods such as the removal of non-semantic high-magnitude directions or the z-scaling of embeddings to expose semantic information in contextualized word embeddings \cite{wolfe2022vast,timkey2021all}; however, such methods necessitate the loss of information, even if that information is syntactic or otherwise non-semantic in nature. The present research introduces a method for assessing both semantic properties and bias in contextualized word embeddings with no postprocessing or loss of information.

\subsection{Bias in Word Embeddings}

Principled and generalizable evaluation of bias in word embeddings is grounded in cognitive psychology literature \cite{greenwald1998measuring,Caliskan_2017}. These foundations, and the word embedding bias tests arising from them, are reviewed below.

\subsubsection{Psychological Foundations for Measuring Machine Bias}
Psychologists quantify the emotional association of a visual or linguistic stimulus using three primary dimensions of affect \cite{kar23659}: valence (pleasantness vs. unpleasantness), arousal (excitement vs. calm), and dominance (control vs. subordination) \cite{RefWorks:RefID:230-osgood1957measurement, RefWorks:RefID:228-mehrabian1974approach,  RefWorks:RefID:229-tellegen1985structures}. Social psychologists have compiled large lexica of affective norms, which reflect widely shared attitudes of human subjects who rate words based on valence, arousal, and dominance \cite{Bradley1999AffectiveNF,warriner2013norms,RefWorks:doc:613a1f068f08743fdb7624b5,vad-acl2018}. A concrete example of a valence norm is that the word “vomit" triggers an unpleasant feeling for most English language speakers, while the word “love" triggers a pleasant feeling.

Valence is the principal dimension of affect that exhibits the strongest affective signal in language \cite{toney2020valnorm}. Psychologists use valence associations to evaluate biased attitudes towards social groups and concepts. \citet{greenwald1998measuring} introduce the Implicit Association Test (IAT), which demonstrated the presence of implicit racial bias favoring European Americans over African Americans by showing that human subjects more readily paired European American names with pleasant words than they did African American names. The IAT inspired the design of the Word Embedding Association Test (WEAT) of \citet{Caliskan_2017}, which demonstrated that a similar phenomenon occurs in static word embeddings, wherein names of European Americans are more similar to pleasant words based on measurements of cosine similarity than are names of African Americans.

In addition to its empirical grounding in social psychology, the WEAT offers theoretical benefits arising from its design as a statistical test: first, the WEAT returns an effect size, Cohen's $d$ \cite{Caliskan_2017}. Cohen's $d$ is defined such that $0.20$ is small, $0.50$ is medium, and $0.80$ is large, and in most cases $d$ ranges between $-2$ and $2$; second, the WEAT returns a $p$-value based on a permutation test \cite{Caliskan_2017}. These qualities make the WEAT a useful method for interpreting the magnitude and statistical significance of bias in embedded representations. While \citet{Caliskan_2017} define the WEAT using cosine similarity, there is no inherent reason that cosine similarity should be the only measurement available for assessing the association of an embedding with some target. For example, \citet{kurita2019measuring} develop a version of the WEAT which uses the masked word prediction objective of BERT to measure differential biases in masked language models.

The WEAT has been adapted previously to study biases in contextualized word embeddings and sentence embeddings formed by language models. \citet{may-etal-2019-measuring} apply the WEAT to measure sentence-level biases in language models such as ELMo and BERT, while \citet{DBLP:conf/nips/TanC19} use a combination of the WEAT as well as method of \citet{may-etal-2019-measuring} to measure biases in a variety of language models such as BERT and GPT-2 \cite{radford2019language}. \citet{10.1145/3461702.3462536} model contextualization as a random effect to measure the overall magnitude of bias across contexts in contextualizing language models. \citet{wolfe2022vast} show that biases exist in the contextualized word embeddings formed by GPT-2 after non-semantic principal components are removed from the embeddings.

\subsubsection{Valence-Based Intrinsic Evaluation of Word Embeddings}

Prior work shows that the correspondence between the human-rated valence of a word and the valence association of its static \cite{toney2020valnorm} or contextualized \cite{wolfe2021low} word embedding can be used to evaluate the intrinsic quality of embedding spaces, and to identify when the geometry of an embedding space interferes with the measurement of semantics using techniques based on cosine similarity \cite{wolfe2022vast}. \citet{wolfe2022vast} find that contextualized word embeddings produced by language models most strongly encode the valence dimension of affect, and that human ratings of dominance also correlate moderately with dominance associations in the contextualized embedding space; arousal, on the other hand, correlates only weakly, with correlations $\rho < 0.30$. This research measures bias in language models using the valence dimension of affect, which corresponds to evaluating biased attitudes towards concepts and social groups.

\subsection{Subspace Projection for Bias Detection and Mitigation}

Another strand of prior work measures bias in word embeddings by identifying a bias subspace. Using $10$ pairs of female-male difference vectors such as "woman" - "man" and "girl" - "boy," \citet{10.5555/3157382.3157584} capture a "gender dimension" in static word embeddings by applying PCA  to the vector differences and finding the component that best accounts for the variance \cite{alma9924548563604107}. Obtaining the projection of other embedded representations of words with this bias subspace yields a metric for quantifying gender bias. \citet{10.5555/3157382.3157584} demonstrate that traditionally masculine occupations such as doctor and pilot project towards masculinity, while traditionally feminine occupations such as nurse and librarian project towards femininity on the gender subspace. Similarly, using difference vectors such as "rich"- "poor," \citet{doi:10.1177/0003122419877135} find the "affluence dimension" in a study of social class in diachronic (chronologically ordered) static word embeddings.

Subspace projection methods have also been adapted to contextualized word embeddings. \citet{zhao-etal-2019-gender} measure and mitigate biases in ELMo’s contextualized word embeddings, and show that a coreference resolution system in ELMo inherits its gender bias. \citet{liang-etal-2020-towards} use a variation of a subspace projection method to measure and mitigate biases in ELMo and BERT’s sentence representations. \citet{DBLP:conf/acl/RavfogelEGTG20} use an iterative variation of a subspace projection method to mitigate biases in contextualized word embeddings, and \citet{DBLP:journals/corr/abs-1904-08783} apply the subspace projection as well as the method of \citet{GONEN19} to measure gender bias in ELMo embeddings. When subspace projection approaches are used to develop techniques for bias mitigation, the success of these interventions is sometimes evaluated using the WEAT \cite{DBLP:conf/acl/RavfogelEGTG20}. 
The present research builds upon prior work by introducing a machine learning method to learn a semantic subspace in the highly contextual and anisotropic upper layers of language models, and introducing a principled statistical test for measuring biases, in an intersectional setting, arising from contextualization in language models.

\section{Data}
\label{sec:data}
The present research examines semantics and bias in five language models based on the transformer architecture of \citet{vaswani2017attention}, which employs a self-attention mechanism to allow word representations to draw information from the representations in the context around them. Models are selected to represent the state-of-the-art for three widely used transformer architectures: decoder-only causal language models; autoencoders; and encoder-decoder models.

\subsection{Language Models}

\noindent\textbf{GPT-Neo} is an open source replication of GPT-3 \cite{NEURIPS2020_1457c0d6}, trained on the next-word prediction objective and employ masked self-attention such that the current token only has access to information from words which precede it in a sentence. GPT-Neo is trained on the Pile, an $825$ GB dataset of English text composed of $22$ diverse and high quality sub-datasets \cite{gao2020pile}. Models trained on the Pile have been shown to outperform models trained on both raw and filtered versions of the Common Crawl on many benchmarks and downstream evaluations \cite{gao2020pile}. Prior work finds that GPT-Neo most strongly encodes human judgments of valence compared to six other language models, including GPT-2 \cite{radford2019language}, T5, and BERT \cite{wolfe2022vast}. This research studies the contextualized word embeddings generated by the 24-layer, 1.3 billion parameter version of GPT-Neo \cite{gpt-neo}. While GPT-Neo is one of the largest and empirically best-performing language models available open source \cite{gao2020pile}, it is still much smaller than the largest version of GPT-3, which has 175 billion parameters \cite{NEURIPS2020_1457c0d6}.

\noindent\textbf{XLNet} is a causal language model that learns bidirectional contexts by permuting the factorization order of text input \cite{NEURIPS2019_dc6a7e65}. XLNet is trained on five corpora:  English Wikipedia, BookCorpus \cite{zhu2015aligning}, Giga5 \cite{parker2011english}, filtered versions of ClueWeb 2012-B \cite{callan2009clueweb09}, and the Common Crawl corpus \cite{buck2014n}. The 12-layer base-cased version is used in this research.

\noindent\textbf{RoBERTa} is an optimized version of the bidirectional "BERT" autoencoder architecture of \citet{devlin-etal-2019-bert}, trained on masked language modeling (prediction of a hidden word) with dynamic masking to prevent memorization of the training data \cite{DBLP:journals/corr/abs-1907-11692}.  RoBERTa is trained on five corpora: English Wikipedia, BookCorpus \cite{zhu2015aligning}, a curated subset of CommonCrawl News \cite{DBLP:journals/corr/abs-1907-11692}, OpenWebText \cite{Gokaslan2019OpenWeb}, and Stories \cite{trinh2018simple}. The 12-layer base version is studied in the present research.

\noindent\textbf{ALBERT} is a parameter-reduced version of the BERT architecture which introduces factorized embedding parameterization, cross-layer parameter sharing, and inter-sentence coherence loss, and outperforms BERT and RoBERTa on a variety of NLP benchmark evaluations \cite{lan2019albert}. ALBERT trains on English Wikipedia and the BookCorpus \cite{zhu2015aligning}. This research uses the 12-layer V-2 base version of ALBERT, which is also trained on additional corpora used to train RoBERTa and XLNet \cite{lan2019albert}.

\noindent\textbf{T5} is an encoder-decoder transformer model that takes text as input and produces text as output, and is trained on a variety of supervised and unsupervised NLP objectives \cite{JMLR:v21:20-074}. T5 is trained on the Colossal Cleaned Common Crawl (C4), a large filtered version of the Common Crawl \cite{JMLR:v21:20-074}. The present research uses the 12 encoder layers of the base version of T5.

All models used are the PyTorch implementations available via the Transformers library of \citet{wolf2020transformers}.

\subsection{Valence Stimuli} As detailed in section \ref{subsec:valdim}, the present research learns a valence dimension by training a support vector classifier (SVC) to form a maximum margin subspace between groups of pleasant and unpleasant words. In keeping with prior research in contextualized word embeddings \cite{10.1145/3461702.3462536,wolfe2021low,wolfe2022vast}, the groups of pleasant and unpleasant words used to measure valence are the stimuli used to measure social biases in the IAT \cite{greenwald1998measuring} and the WEAT \cite{Caliskan_2017}.

Pleasant vs. Unpleasant stimuli obtained from \citet{Caliskan_2017} to learn an affective valence dimension are included below.

\noindent\textbf{Pleasant:} caress, freedom, health, love, peace, cheer, friend, heaven, loyal, pleasure, diamond, gentle, honest, lucky, rainbow, diploma, gift, honor, miracle, sunrise, family, happy, laughter, paradise, vacation

\noindent\textbf{Unpleasant:} abuse, crash, filth, murder, sickness, accident, death, grief, poison, stink, assault, disaster, hatred, pollute, tragedy, divorce, jail, poverty, ugly, cancer, kill, rotten, vomit, agony, prison

\begin{table*}[h!]
  \caption{Category terms chosen from related work in AI bias and social psychology to represent human social biases. The term $r$ denotes the frequency ratio of the first category to the second according to Google ngrams corpus of English books \cite{lin2012syntactic}.}
  \label{tab:char}
  \begin{tabular}{llr|llr}
    \toprule
    \textbf{Social Bias} & \textbf{Categories} & \textbf{$r$} & \textbf{Social Bias} & \textbf{Categories} & \textbf{$r$}\\
    \midrule
    age & young, old &  $0.59$ & social class & affluent, destitute &  $0.55$\\
    weight & thin, fat &  $1.40$ & race & white, black &  $1.30$ \\
    height & tall, short &  $0.12$ & sexual orientation & heterosexual, homosexual &  $0.64$\\
    intelligence & smart, stupid &  $0.98$ & religion & christian, muslim & $14.36$\\
    education & educated, ignorant &  $1.70$ & gender & cisgender, transgender & $0.05$\\ 
    literacy & literate, illiterate &  $0.85$ & sex & male, female & $0.98$ \\
  \bottomrule
\end{tabular}
\end{table*}

\subsection{Social Biases and Categories} The present research designs a method for language models to study the effects of multiple biases  interacting in a single string input in an intersectional setting. This requires the identification of a variety of societal biases which may overlap and compound each other in contextualizing language models, as they are known to human society. Drawing on prior work in psychology and AI bias \cite{doi:10.1177/0003122419877135,jenkins1958atlas}, $12$ western societal biases are identified for study in this work. These include biases based on age, weight, height, intelligence, education, literacy, social class, race, sexual orientation, religion, gender, and sex. 

For each of these $12$ social biases, two categories are selected such that bias arising from the difference in these categories can be measured. For example, the categories "tall" and "short" are selected to measure bias based on height. Because word frequency can affect the representational quality of a word in a contextualizing language model \cite{wolfe2021low}, categories are selected such that they have relatively balanced frequency based on human usage as measured using Google ngrams \cite{lin2012syntactic}. For example, though "educated" and "uneducated" could be used to quantify biases based on educational attainment, "uneducated" is used roughly $10$ times less frequently than "educated" in ngrams \cite{lin2012syntactic}. To balance the frequency of the words, "ignorant" is selected as the second category in the pair with "educated."

Table~\ref{tab:char} shows the social biases evaluated and their corresponding categories. Column $r$ describes the term frequency ratio of the first category to the second category. Although the intention is to represent each bias with category terms that have similar rates of frequency, the categories for gender bias are highly imbalanced. For the lack of a more suitable alternative, "cisgender" remains one of the gender categories, despite its imbalance with the more commonly used term "transgender."

Many of the biases examined in this research could be represented with more than two categories. There are, for example, more religions, sexual orientations, and genders than those captured here. This research introduces a new method and demonstrates that it captures these well-studied social biases. The method generalizes beyond the categories defined herein.

\section{Approach}
\label{sec:approach}

\begin{figure}
    \centering
    \includegraphics[scale = 0.12]{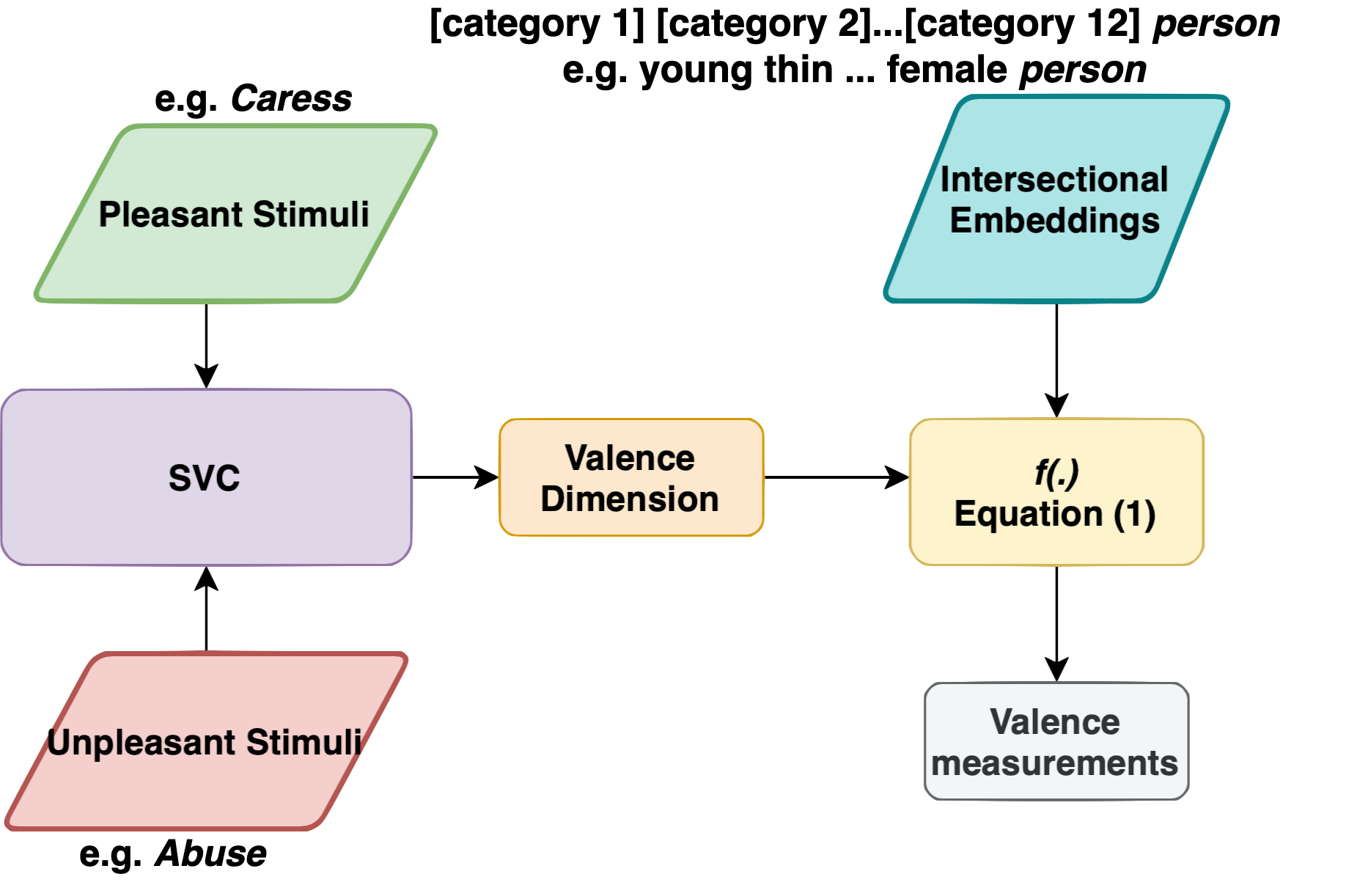}
    \caption{A support vector classifier is used to learn the valence dimension in the upper layers of contextualizing language models. Biases related to pleasantness are evaluated by taking projection product of the contextualized representation of "person" at the end of a context with the learned valence dimension.}
    \label{fig:approach}
\end{figure}

The present research describes a new method for measuring biases based on valence in contextualized word embeddings. This involves first learning an affective dimension in the contextualized embedding space, and then measuring bias based on the projection product of a contextualized word embedding with the learned dimension. Figure~\ref{fig:approach} summarizes the approach.

\subsection{Learning an Affective Dimension}
\label{subsec:valdim}

While recent work shows that the semantic properties of contextualized word embeddings, including valence \cite{wolfe2022vast}, can be isolated by removing top principal components, these methods have the significant drawback of postprocessing the embeddings, and removing information from the model's representations. To mitigate this constraint, the present research proposes a method which requires no postprocessing of the embedding space, but instead learns a property of the space against which contextualized representations can be measured.

The valence direction is learned in the contextualized embedding space by training an SVC with a linear kernel given the high dimensionality of the space. For valence, the SVC is trained to classify contextualized representations of $25$ pleasant words and $25$ unpleasant words such that the separating subspace between the pleasant words and the unpleasant words maximizes the distance between them. The coefficients of the separating subspace are extracted, and used as a valence dimension of the contextualized embedding space, respectively. The words used to learn a separating subspace are input to the model in the decontextualized setting, \textit{i.e.,} with no surrounding context. Each decontextualized word is preceded by the <BOS> (Beginning of Sequence) token, extracted from the model tokenizer.

\subsection{Contextualizations of "Person"}
\label{subsec: construction}

While prior work evaluates word embedding bias on the word level \cite{Caliskan_2017,wolfe2022vast} or the sentence level \cite{may-etal-2019-measuring}, this research measures biases resulting from the contextualization of the word "person," such that it is altered along the valence dimension. More concretely, the question under consideration is whether the word person becomes more pleasant or unpleasant when it occurs in a sentence with a word like "transgender" or "cisgender" (\textit{e.g.,} "a transgender person"). Because causal language models employ masked self-attention such that the current word only has access to the information of the words which precede it, this research positions the word person at the end of the sentence, such that information can be drawn from all other words in the sentence. Models such as BERT and T5, which employ bidirectional self-attention, are also able to retrieve information from any word in a context given this format.

\subsection{Measuring Valence Associations}
\label{subsec: valmezh}

The valence association of a word's embedded representation is measured by its orthogonal scalar projection onto the learned affective dimension. For a vector $v$, and subspace $U$ defined by $n$ orthogonal vectors $u_1, u_2,\dots,u_n$, the scalar projection of $v$ onto $U$ is computed as follows:

\begin{equation}
    \textrm{S}(v,U) = \sum_{i=1}^{n}\frac{(v.u_i)}{(u_i.u_i)}
    \label{eq:proj}
\end{equation}
\\
\noindent where $(a.b)$ refers to the dot product of $a$ and $b$. The valence dimension is learned such that positive values of $s$ correspond to greater association with pleasantness (\textit{i.e.,} high-valence words will project onto the positive side of the valence dimension), while negative values of $s$ correspond to greater association with unpleasantness (projection onto the negative side of the valence dimension). 

\subsection{Quantifying Differential Bias Using the SC-WEAT}
The WEAT and the SC-WEAT measure biased associations and return two values: an effect size, Cohen's $d$, and a $p$-value based on a permutation test \cite{Caliskan_2017}. \citet{Caliskan_2017} define the WEAT as using cosine similarity as a means of assessing the similarity between two embedded representations, as this distance metric reflects a widespread paradigm for measuring similarity in static word embeddings \cite{Caliskan_2017,mikolov-etal-2013-linguistic}. However, the WEAT is a statistical method for assessing differential similarity of two sets of targets (\textit{e.g.,} two social groups) with two sets of attributes (\textit{e.g.,} pleasantness and unpleasantness), and is not necessarily dependent upon cosine similarity as a distance metric when a more appropriate measure is validated for an embedding space. For the present research, a WEAT is defined to capture the differential bias of two words in contextualizing language models, based on their projection product with the valence dimension. The formula of the SC-WEAT is readily adaptable for this purpose, as it measures the differential similarity of a single target vector with two attribute groups: 

\begin{equation}
\frac{\textrm{mean}_{a\in A}\textrm{S}(a,U) - \textrm{mean}_{b\in B}\textrm{S}(b,U)}{\textrm{std\_dev}_{x \in A \cup B}\textrm{S}(x,U)}
\end{equation}
\\
In this case, the learned affective dimension $U$ (\textit{i.e.,} valence) is used as the target. To measure the differential bias for two words across contexts, the $A$ attribute group is defined to include the embedded representations $a$ of the word "person" in all of the sentences which include a certain attribute word, such as "transgender," and a $B$ attribute group is defined to include a set of sentences which are identical to the $A$ group, but with the target word replaced with an opposing category word, such as "cisgender," for which the differential bias effect size will be obtained. The bias measurement is defined as the difference in the mean projection product of the $A$ group with the valence dimension and the $B$ group with the valence dimension, divided by the joint standard deviation of projection products, commensurate with Cohen's $d$. A p-value is obtained using the same permutation test as employed in the SC-WEAT \cite{Caliskan_2017}.

\section{Experiments}

In this section, details are provided for three different experiments and their results. Experiment 1 examines the utility of the learned valence dimension for capturing semantics in language models. Experiment 2 studies differential biases based on valence in language models. Experiment 3 examines the words most biased based on association with valence.

\subsection{Evaluating Learned Affective Dimensions Against Human Judgments of Semantics}
\label{sec:eval-learn}

The utility of the learned dimension for representing valence in the contextualized word embedding space is assessed using the ValNorm method of \citet{toney2020valnorm}. ValNorm is an intrinsic evaluation task that obtains the correlation (Pearson's $\rho$) of a word's human valence rating in a valence lexicon with the SC-WEAT valence association of its embedded representation. \citet{toney2020valnorm} employ three valence lexica in evaluating ValNorm, of which we select Bellezza's lexicon \cite{RefWorks:doc:613a1f068f08743fdb7624b5}, a set of $399$ words rated by human subjects based on pleasantness, which \citet{wolfe2022vast} show is sensitive to the presence of non-semantic high-magnitude directions in language models. To show that the method is effective in the highly contextual upper layers of language models \cite{ethayarajh-2019-contextual}, a ValNorm score (Pearson's $\rho$) is obtained at every layer of the language model using the projection product with the valence subspace as a word's valence association in the model.

\begin{figure*}[htbp]
\begin{tikzpicture}
\begin{axis} [
    height=4cm,
    width=17.5cm,
    line width = .5pt,
    ymin = 0, 
    ymax = 1,
    xmin=-.5,
    xmax=24.5,
    ylabel=Pearson's $\rho$,
    ylabel shift=-5pt,
    ylabel near ticks,
    xtick = {0,1,2,3,4,5,6,7,8,9,10,11,12,13,14,15,16,17,18,19,20,21,22,23,24},
    xtick pos=left,
    ytick pos = left,
    title=GPT-Neo 1.3B ValNorm Score,
    xlabel= {Layer},
    legend style={at={(.35,0.13)},anchor=south west,nodes={scale=.8, transform shape}}
]

\addplot [thick,dotted,mark=triangle*,color=black] coordinates {(0, 0.31124942548239776) (1, 0.45856192776438337) (2, 0.46099095578531907) (3, 0.5076276618550857) (4, 0.5215707395610084) (5, 0.6209068922237655) (6, 0.661180845360382) (7, 0.7477194661606595) (8, 0.7835149510368018) (9, 0.8255847873759342) (10, 0.8317788631631186) (11, 0.8441742525597913) (12, 0.8481714707194792) (13, 0.8574512731560358) (14, 0.8584373814101101) (15, 0.86461140203056) (16, 0.8689867164225007) (17, 0.867289478075385) (18, 0.8629965509515778) (19, 0.8586587849103062) (20, 0.8565884556002018) (21, 0.8510245552906203) (22, 0.8459418096984617) (23, 0.829682686555315) (24, 0.80581433958729)};

\addplot [thick,solid,mark=*,color=red] coordinates {(0, 0.22832317959253443) (1, 0.45430068115401295) (2, 0.47807172925637875) (3, 0.5238138835892544) (4, 0.5313785568153062) (5, 0.5870079519655985) (6, 0.6300680107018147) (7, 0.7387675094286589) (8, 0.7647479397302376) (9, 0.8086104722590517) (10, 0.8053742191936553) (11, 0.8226707289734556) (12, 0.8246484798644499) (13, 0.8243999528767771) (14, 0.8188944711350835) (15, 0.8197854639315864) (16, 0.8129064228903367) (17, 0.8183745037136569) (18, 0.808834526046765) (19, 0.7984285076028579) (20, 0.8001591716175176) (21, 0.7866139108298392) (22, 0.7764577139397527) (23, 0.7005430537773167) (24, 0.5596788000419778)};
\vspace{5mm}
\legend {{\small SVC ValNorm,\small Traditional ValNorm}};
\end{axis}
\end{tikzpicture}
\begin{tikzpicture}
\begin{axis} [
    height=4cm,
    width=9cm,
    line width = .5pt,
    ymin = 0, 
    ymax = 1,
    xmin=-.5,
    xmax=12.5,
    ylabel=Pearson's $\rho$,
    ylabel shift=-5pt,
    ylabel near ticks,
    xtick = {0,1,2,3,4,5,6,7,8,9,10,11,12},
    xtick pos=left,
    ytick pos = left,
    title=ALBERT ValNorm Score,
    xlabel= {Layer},
    legend style={at={(.01,0.05)},anchor=south west,nodes={scale=.8, transform shape}}
]

\addplot [thick,dotted,mark=triangle*,color=black] coordinates {(0, 0.8866016648726049) (1, 0.8864236006013012) (2, 0.8758118717056897) (3, 0.8778536279951458) (4, 0.8601288229595245) (5, 0.8209423301538217) (6, 0.8059032640074818) (7, 0.7647027563919806) (8, 0.7484048567919372) (9, 0.7335523000331383) (10, 0.7249859258245324) (11, 0.7201825043011637) (12, 0.7305607862967737)};

\addplot [thick,solid,mark=*,color=red] coordinates {(0, 0.8515598226756844) (1, 0.8527279412417801) (2, 0.825423584454182) (3, 0.763916475656785) (4, 0.6822732134633943) (5, 0.5399573347467373) (6, 0.5121494320520439) (7, 0.37097694086362637) (8, 0.3918226473328055) (9, 0.41927446785325556) (10, 0.47616607242938974) (11, 0.449079976289433) (12, 0.5254781798256596)};

\end{axis}
\end{tikzpicture}
\begin{tikzpicture}
\begin{axis} [
    height=4cm,
    width=9cm,
    line width = .5pt,
    ymin = 0, 
    ymax = 1,
    xmin=-.5,
    xmax=12.5,
    ylabel=Pearson's $\rho$,
    ylabel near ticks,
    ylabel shift=-5pt,
    xtick = {0,1,2,3,4,5,6,7,8,9,10,11,12},
    xtick pos=left,
    ytick pos = left,
    title=T5 ValNorm Score,
    xlabel= {Layer},
    legend style={at={(.29,0.13)},anchor=south west,nodes={scale=.8, transform shape}}
]

\addplot [thick,dotted,mark=triangle*,color=black] coordinates {(0, 0.8045485348425527) (1, 0.7928616006431589) (2, 0.8035766843526571) (3, 0.8334875204423082) (4, 0.8630991060438888) (5, 0.8552359574448781) (6, 0.8452388001234292) (7, 0.8417884191335342) (8, 0.8388832945282049) (9, 0.8421314482366287) (10, 0.8341797082003671) (11, 0.8180090142785397) (12, 0.8317107480450597)};

\addplot [thick,solid,mark=*,color=red] coordinates {(0, 0.7718901284991386) (1, 0.7923703755201227) (2, 0.8029655444519321) (3, 0.828226140343298) (4, 0.8453945473229637) (5, 0.8129453244054927) (6, 0.7888697734216339) (7, 0.7656584578727599) (8, 0.7655278506387102) (9, 0.7508352562071481) (10, 0.7294148593394896) (11, 0.7063189332646396) (12, 0.7765626847666387)};

\end{axis}
\end{tikzpicture}
\begin{tikzpicture}
\begin{axis} [
    height=4cm,
    width=9cm,
    line width = .5pt,
    ymin = 0, 
    ymax = 1,
    xmin=-.5,
    xmax=12.5,
    ylabel=Pearson's $\rho$,
    ylabel near ticks,
    ylabel shift=-5pt,
    xtick = {0,1,2,3,4,5,6,7,8,9,10,11,12},
    xtick pos=left,
    ytick pos = left,
    title=RoBERTa ValNorm Score,
    xlabel= {Layer},
    legend style={at={(.17,0.01)},anchor=south west,nodes={scale=.8, transform shape}}
]

\addplot [thick,dotted,mark=triangle*,color=black] coordinates {(0, 0.41027643805036423) (1, 0.45898136634587344) (2, 0.519621327875848) (3, 0.628801821191097) (4, 0.6551297786971798) (5, 0.6418574503118906) (6, 0.6142312596948465) (7, 0.5592028493048333) (8, 0.560573408973442) (9, 0.5531520886166109) (10, 0.5569679366454174) (11, 0.5465741862383677) (12, 0.47524806375308976)};

\addplot [thick,solid,mark=*,color=red] coordinates {(0, 0.40626792068224654) (1, 0.4151130375440322) (2, 0.44744255123976) (3, 0.5262839266464963) (4, 0.5578372716639537) (5, 0.5764288566287595) (6, 0.5437624614591923) (7, 0.54538401745268) (8, 0.5103830546550749) (9, 0.501887517859746) (10, 0.4709590132015168) (11, 0.4182866751283137) (12, 0.3282398485945318)};

\end{axis}
\end{tikzpicture}
\begin{tikzpicture}
\begin{axis} [
    height=4cm,
    width=9cm,
    line width = .5pt,
    ymin = 0, 
    ymax = 1,
    xmin=-.5,
    xmax=12.5,
    ylabel=Pearson's $\rho$,
    ylabel near ticks,
    ylabel shift=-5pt,
    xtick = {0,1,2,3,4,5,6,7,8,9,10,11,12},
    xtick pos=left,
    ytick pos = left,
    title=XLNet ValNorm Score,
    xlabel= {Layer},
    legend style={at={(.04,0.05)},anchor=south west,nodes={scale=.8, transform shape}}
]

\addplot [thick,dotted,mark=triangle*,color=black] coordinates {(0, 0.7988896830266077) (1, 0.7916189233448148) (2, 0.801378172698569) (3, 0.8073211992922658) (4, 0.8177567537512314) (5, 0.8092878654936506) (6, 0.8156254539129817) (7, 0.7969675467820313) (8, 0.785821592993643) (9, 0.8071642802033446) (10, 0.7689834647422983) (11, 0.7417357527588113) (12, 0.7376417640302859)};

\addplot [thick,solid,mark=*,color=red] coordinates {(0, 0.7433089879749093) (1, 0.7186194699884015) (2, 0.733610642633279) (3, 0.7293606353675545) (4, 0.6668160938451164) (5, 0.6522771556354325) (6, 0.6253237901681626) (7, 0.559047031117656) (8, 0.4694546477065341) (9, 0.46569806510642514) (10, 0.5315060707185097) (11, 0.358562788319786) (12, 0.3947956343238502)};

\end{axis}
\end{tikzpicture}

\caption{Across five contextualizing language models, using a support vector classifier to learn the valence dimension improves ValNorm evaluation scores in the upper layers of the language model over comparable results obtained using cosine similarity, without the need for postprocessing of embeddings. This result suggests the robustness of the methods proposed in this research for capturing semantics across widely varying language modeling architectures and pretraining objectives.}
\label{fig:gpt2_valnorm}
\end{figure*}
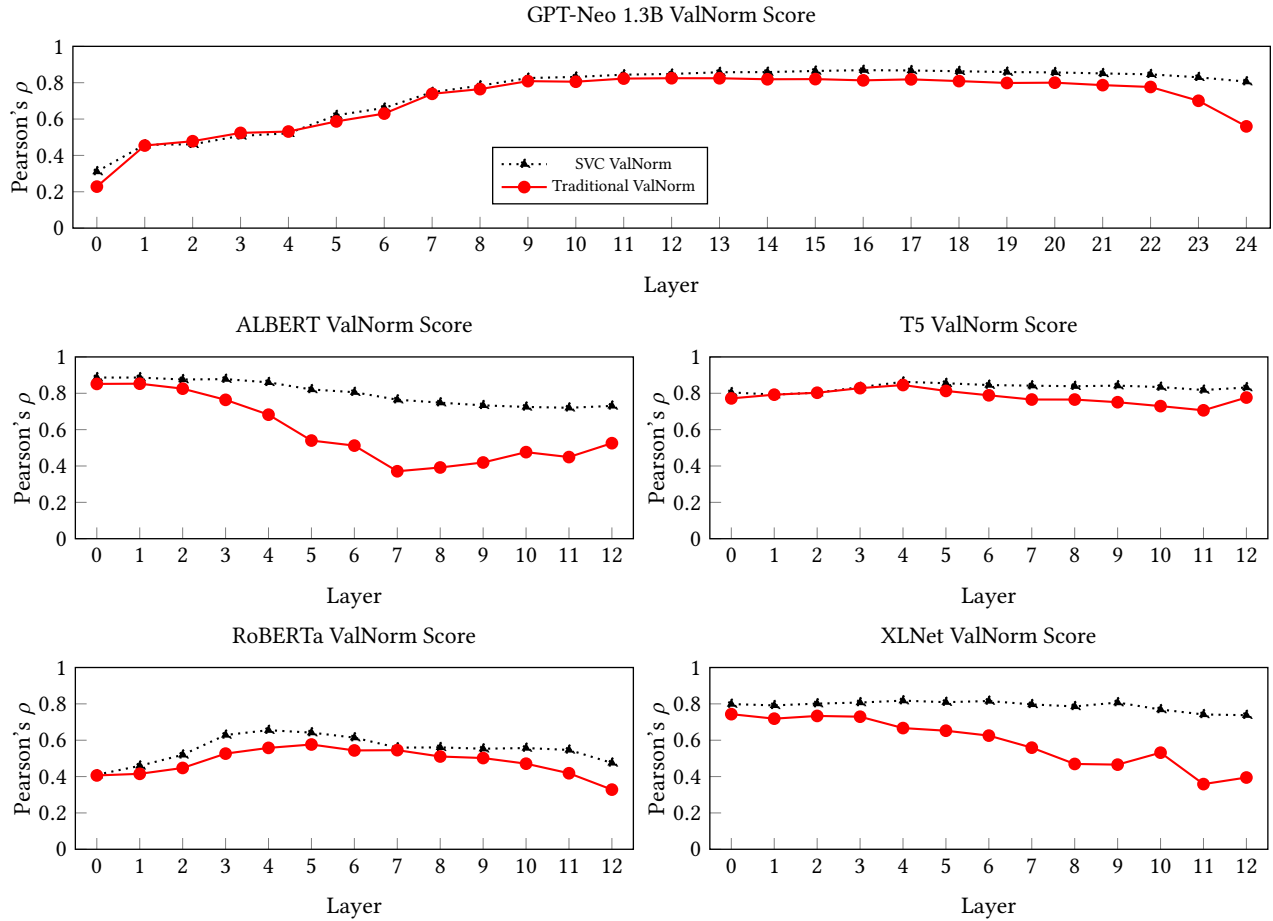

\subsection{Bias Evaluation Using SC-WEAT} 
\label{sec:ex1}

The categories derived from the $12$ social biases considered in this research are placed into a sentence in the order shown in Table \ref{tab:char}, \textit{i.e.,} "a young thin [...] female person." While maintaining the order of the biases, the context is altered such that every category occurs in a sentence with every other combination of categories, except for its own opposing category. This leads to a total of $2^{12}$, or $4,096$ contexts. Each category occurs in exactly half of these contexts, or $2,048$ occurrences.

The order of the social bias categories in the sentence template was chosen in an attempt to make the sentences sound more natural. For example, "thin female person" is used more frequently than "female thin person" \cite{lin2012syntactic}. Ideally, one should generate all permutations of the social biases to eliminate the impact of word order on the bias captured by "person" at the end of the sentence. However, the total number of permutations in this experiment would have been about $2$ trillion sentences which was beyond our computational capacity. Future work can investigate the impact of word order on bias computations.

The two categories described in Table \ref{tab:char} are selected for each of the $12$ biases examined in this research, with the first category (\textit{e.g.,} affluent) set as the $A$ attribute, and the second category (\textit{e.g.,} destitute) set as the $B$ attribute, such that a positive effect size reflects stereotype-congruent bias (\textit{e.g.,} affluent individuals are evaluated more positively than destitute individuals). For each category, the $2,048$ sentence combinations in which the $A$ attribute occurs are selected, and the embedded representation for the word "person" is obtained for each of these contexts. The same process is repeated for the $B$ attribute, and the two sets of embeddings are used as input to the projection product SC-WEAT. To obtain the most contextual representation produced by a transformer, \textit{i.e.,} the representation most altered by the words in its context, the contextualized word embedding in the top (output) layer of the model is obtained, commensurate with prior research which finds that top layers of language models are the most contextual \cite{ethayarajh-2019-contextual,wolfe2022vast}. A bias effect size $d$ and a $p$-value are obtained for each test. In total, five transformers of varying architectures and pretraining objectives are examined.

\subsection{Identifying the Strongest Biases Across Contexts}
\label{subsec:order}

A final experiment examines the most biased categories in GPT-Neo, the largest of the language models studied herein. We choose GPT-Neo because \citet{nadeem2020stereoset} observes that larger, better-performing language models are also more biased. Five historically disadvantaging societal biases are selected for study: race, sex, religion, gender, and sexual orientation. The ten categories associated with these concepts are drawn from Table \ref{tab:char}. Sentences are created with five categories present per sentence (\textit{e.g.,} a white female cisgender heterosexual Christian person). All possible permutations are generated using the ten categories in question, such that every category is seen in combination with every other category in every position in the sentence. The total number of permutation of phrases constructed in this manner is $3,840$. The valence projection product is obtained for the embedded representation of the word "person" in every generated sentence. The characteristics of the top $10\%$ most positively valenced and top $10\%$ most negatively valenced five-category sentences are examined. For these subsets of the generated sentences, the percentage of the time each category word occurs in each of the positions preceding "person" is quantified.

\section{Results}

The evidence indicates that learning the valence is useful for detecting semantics and social biases in the contextual and anisotropic upper layers of language models.

\begin{table*}[htbp]
\centering
{
\begin{tabular}
{|l|c|l|c|l|c|l|c|l|c|l|}
 \hline
 \multicolumn{11}{|c|}{SC-WEAT Differential Valence Association} \\
 \hline
 \multirow{2}{*}{Bias Test} & \multicolumn{2}{c}{ALBERT} &  \multicolumn{2}{|c|}{GPT-Neo} & \multicolumn{2}{|c|}{RoBERTa} & \multicolumn{2}{|c|}{T5} & \multicolumn{2}{|c|}{XLNet}\\
 \cline{2-11}
 &\multicolumn{1}{|c|}{$d$} & \multicolumn{1}{|c|}{$p < $} &  \multicolumn{1}{|c|}{$d$} & \multicolumn{1}{|c|}{$p < $} & 
\multicolumn{1}{|c|}{$d$} & \multicolumn{1}{|c|}{$p < $} & 
\multicolumn{1}{|c|}{$d$} & \multicolumn{1}{|c|}{$p < $} & 
\multicolumn{1}{|c|}{$d$} & \multicolumn{1}{|c|}{$p < $} \\
 \hline

 young vs. old & \cellcolor{gray!0}$-0.68$  & $10^{-30}$ & \cellcolor{gray!50.0}$0.50$  & $10^{-30}$ & \cellcolor{gray!33.0}$0.33$  & $10^{-30}$ & \cellcolor{gray!30.0}$0.30$  & $10^{-30}$ & \cellcolor{gray!0}$-0.67$  & $10^{-30}$ \\ 
thin vs. fat & \cellcolor{gray!0}$-0.02$  & $n.s.$ & \cellcolor{gray!56.00000000000001}$0.56$  & $10^{-30}$ & \cellcolor{gray!20.0}$0.20$  & $10^{-9}$ & \cellcolor{gray!0}$-0.13$  & $10^{-4}$ & \cellcolor{gray!0}$-0.06$  & $.05$ \\ 
tall vs. short & \cellcolor{gray!22.0}$0.22$  & $10^{-12}$ & \cellcolor{gray!0}$-0.06$  & $.05$ & \cellcolor{gray!20.0}$0.20$  & $10^{-9}$ & \cellcolor{gray!0}$-0.27$  & $10^{-16}$ & \cellcolor{gray!86.0}$0.86$  & $10^{-30}$ \\ 
smart vs. stupid & \cellcolor{gray!2.0}$0.02$  & $n.s.$ & \cellcolor{gray!56.00000000000001}$0.56$  & $10^{-30}$ & \cellcolor{gray!82.0}$0.82$  & $10^{-30}$ & \cellcolor{gray!0}$-0.01$  & $n.s.$ & \cellcolor{gray!48.0}$0.48$  & $10^{-30}$ \\ 
educated vs. ignorant & \cellcolor{gray!32.0}$0.32$  & $10^{-30}$ & \cellcolor{gray!92.0}$0.92$  & $10^{-30}$ & \cellcolor{gray!81.0}$0.81$  & $10^{-30}$ & \cellcolor{gray!0}$-0.22$  & $10^{-12}$ & \cellcolor{gray!0}$-0.04$  & $n.s.$ \\ 
literate vs. illiterate & \cellcolor{gray!0}$-0.18$  & $10^{-10}$ & \cellcolor{gray!17.0}$0.17$  & $10^{-9}$ & \cellcolor{gray!0}$-0.05$  & $.05$ & \cellcolor{gray!1.0}$0.01$  & $n.s.$ & \cellcolor{gray!11.0}$0.11$  & $10^{-4}$ \\ 
affluent vs. destitute & \cellcolor{gray!67.0}$0.67$  & $10^{-30}$ & \cellcolor{gray!100}$1.10$  & $10^{-30}$ & \cellcolor{gray!12.0}$0.12$  & $10^{-3}$ & \cellcolor{gray!0}$-0.03$  & $n.s.$ & \cellcolor{gray!52.0}$0.52$  & $10^{-30}$ \\ 
white vs. black & \cellcolor{gray!35.0}$0.35$  & $10^{-30}$ & \cellcolor{gray!0}$-0.12$  & $10^{-3}$ & \cellcolor{gray!14.000000000000002}$0.14$  & $10^{-5}$ & \cellcolor{gray!31.0}$0.31$  & $10^{-30}$ & \cellcolor{gray!0}$-0.08$  & $.01$ \\ 
heterosexual vs. homosexual & \cellcolor{gray!35.0}$0.35$  & $10^{-30}$ & \cellcolor{gray!64.0}$0.64$  & $10^{-30}$ & \cellcolor{gray!12.0}$0.12$  & $10^{-4}$ & \cellcolor{gray!61.0}$0.61$  & $10^{-30}$ & \cellcolor{gray!40.0}$0.40$  & $10^{-30}$ \\ 
christian vs. muslim & \cellcolor{gray!27.0}$0.27$  & $10^{-30}$ & \cellcolor{gray!0}$-0.15$  & $10^{-6}$ & \cellcolor{gray!0}$-0.63$  & $10^{-30}$ & \cellcolor{gray!1.0}$0.01$  & $n.s.$ & \cellcolor{gray!0}$-0.16$  & $10^{-6}$ \\ 
cisgender vs. transgender & \cellcolor{gray!100}$1.34$  & $10^{-30}$ & \cellcolor{gray!24.0}$0.24$  & $10^{-14}$ & \cellcolor{gray!100}$1.22$  & $10^{-30}$ & \cellcolor{gray!9.0}$0.09$  & $.01$ & \cellcolor{gray!12.0}$0.12$  & $10^{-4}$ \\ 
male vs. female & \cellcolor{gray!27.0}$0.27$  & $10^{-30}$ & \cellcolor{gray!10.0}$0.10$  & $10^{-3}$ & \cellcolor{gray!10.0}$0.10$  & $10^{-3}$ & \cellcolor{gray!0}$-0.93$  & $10^{-30}$ & \cellcolor{gray!1.0}$0.01$  & $n.s.$ \\

 \hline
\end{tabular}
\vspace{3mm}
\caption{Across five language model architectures, the most severe biases occur for sexual orientation and gender identity, with positive effect sizes obtained from all five models assessed. GPT-Neo includes six effect sizes of $.5$ or greater, the largest number of any language model, corresponding to the observation of \citet{nadeem2020stereoset} that larger, better-performing language models are also more biased.}
\label{tab:valence_biases}}
\end{table*}

\begin{figure*}
\begin{minipage}[c]{0.6\textwidth}
\resizebox{10cm}{!}{%

\begin{tikzpicture}
[
dot/.style = {circle, fill=purple,inner sep=0pt, minimum size=12pt},
every label/.append style = {inner sep=2pt,color = black, font= \fontsize{28}{0}\bfseries},
    thick      
, 
dot2/.style = {fill=blue,inner sep=0pt, minimum size=12pt},
                        ]

\newcount\nOne; \nOne=-10
\def\w{18}      
\def\n{29}      
\def\noffset{1} 
\def\nskip{3}   
\def\la{2.00}   
\def\lt{0.20}   
\def\ls{0.15}   

\def\myx(#1){{(#1-\nOne)*\w/\n}}
\def\arrowLabel(#1,#2,#3,#4,#5){
\def\xy{(#1-\nOne)*\w/\n}; \pgfmathparse{int(#2*100)};
\ifnum \pgfmathresult<0
  \def\yyp{{(\lt*(-0.10+#2))}}; \def\yyw{{(\yyp-\la*\lt*#3)}}
  \draw[<-,thick,black!50!blue,align=center]
    (\myx(#1),\yyp) -- (\myx(#1),\yyw)
    node[below,black!80!blue] {#4}; 
\else
  \def\yyp{{(\lt*(0.10+#2)}}; \def\yyw{{(\yyp+\la*\lt*#3)}}
  \draw[<-,thick,black!50!blue,align=center]
    (#1,\yyp) -- (#1 + #5 ,\yyw)
    node[above,black!80!blue] {#4};
\fi}

\newcommand\vara{17}
\newcommand\varc{4}

\draw [thick,->](0,0) -- + (45,0) node[above] {\fontsize{26}{0}\color{gray}\bfseries{Boundary}};
\draw [thick,->](0,0) -- + (0,4) node[above] {\fontsize{26}{0}\color{gray}\bfseries{Pleasant}};
\draw [thick,->](0,0) -- + (0,-25) node[below] {\fontsize{26}{0}\color{gray}\bfseries{Unpleasant}};
\draw [thick,->](-2,-8) -- + (0,0) node[rotate=90] {\fontsize{34}{0}\color{gray}\bfseries{Projection Onto Valence Dimension}};

\node[dot2,label=below:old] at (1.5+0*\varc,-1.217632569659266*\vara) (old){};
\node[dot,label=above:young] at (1.5+0*\varc,-0.6732039374473352*\vara) (young){};
\draw[-, green] (old.south) -- (young.north);
\node[dot2,label=below:fat] at (1.5+1*\varc,-1.245473699980903*\vara) (fat){};
\node[dot,label=above:thin] at (1.5+1*\varc,-0.6453628071256983*\vara) (thin){};
\draw[-, green] (fat.south) -- (thin.north);
\node[dot2,label=below:short] at (1.5+2*\varc,-0.9537315287003328*\vara) (short){};
\node[dot,label=above:tall] at (1.5+2*\varc,-0.9371049784062685*\vara) (tall){};
\draw[-, green] (short.south) -- (tall.north);
\node[dot2,label=below:stupid] at (1.5+3*\varc,-1.2423370761329942*\vara) (stupid){};
\node[dot,label=above:smart] at (1.5+3*\varc,-0.6484994309736072*\vara) (smart){};
\draw[-, green] (stupid.south) -- (smart.north);
\node[dot2,label=below:ignorant] at (1.5+4*\varc,-1.414599854261447*\vara) (ignorant){};
\node[dot,label=above:educated] at (1.5+4*\varc,-0.47623665284515426*\vara) (educated){};
\draw[-, green] (ignorant.south) -- (educated.north);
\node[dot2,label=below:illiterate] at (1.5+5*\varc,-1.1294706246802646*\vara) (illiterate){};
\node[dot,label=above:literate] at (1.5+5*\varc,-0.9454182535533007*\vara) (literate){};
\draw[-, green] (illiterate.south) -- (literate.north);
\node[dot2,label=below:destitute] at (1.5+6*\varc,-1.4753283335368848*\vara) (destitute){};
\node[dot,label=above:affluent] at (1.5+6*\varc,-0.4155081735697166*\vara) (affluent){};
\draw[-, green] (destitute.south) -- (affluent.north);
\node[dot2,label=below:white] at (1.5+7*\varc,-0.9977884722472492*\vara) (white){};
\node[dot,label=above:black] at (1.5+7*\varc,-0.8930480348593522*\vara) (black){};
\draw[-, green] (white.south) -- (black.north);
\node[dot2,label=below:homosexual] at (1.5+8*\varc,-1.2509607184538702*\vara) (homosexual){};
\node[dot,label=above:heterosexual] at (1.5+8*\varc,-0.6398757886527313*\vara) (heterosexual){};
\draw[-, green] (homosexual.south) -- (heterosexual.north);
\node[dot2,label=below:christian] at (1.5+9*\varc,-1.0183884349652805*\vara) (christian){};
\node[dot,label=above:muslim] at (1.5+9*\varc,-0.8724480721413208*\vara) (muslim){};
\draw[-, green] (christian.south) -- (muslim.north);
\node[dot2,label=below:transgender] at (1.5+10*\varc,-1.0602365568023746*\vara) (transgender){};
\node[dot,label=above:cisgender] at (1.5+10*\varc,-0.8305999503042267*\vara) (cisgender){};
\draw[-, green] (transgender.south) -- (cisgender.north);
\node[dot2,label=below:female] at (1.5+11*\varc,-1.0378076899202484*\vara) (female){};
\node[dot,label=above:male] at (1.5+11*\varc,-0.9454182535533007*\vara) (male){};
\draw[-, green] (female.south) -- (male.north);

\end{tikzpicture}   
}
\end{minipage}\hfill
\begin{minipage}[c]{0.4\textwidth}
\caption{Differences in the mean valence of the word "person" when it co-occurs with the above categories in $4,096$ phrases. Length of green lines represents the magnitude of differential valence for each pair of categories. Red circles indicate stereotypically higher-valence categories, while red squares represent stereotypically lower-valence categories.}
\label{fig:avg1}
\end{minipage}
\end{figure*}
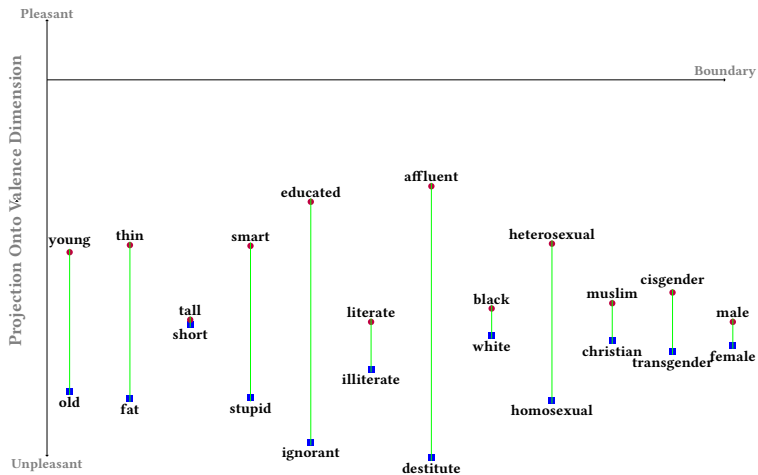

\subsection{Evaluating the Learned Affective Dimension}

Across five state-of-the-art transformer language models with different architectures, tokenization algorithms, and training objectives, learning an affective dimension in the embedding space outperforms cosine similarity on the ValNorm intrinsic evaluation task with no postprocessing of the embeddings. The effect is especially noticeable in the highly contextual upper layers of these models, where non-semantic high-magnitude directions distort measurements of semantics based solely on cosine similarity. As shown in Figure~\ref{fig:gpt2_valnorm}, the ValNorm score (Pearson's $\rho$) drops to $0.56$ in the top layer of GPT-Neo 1.3B when using cosine similarity, but stays high, at $0.81$, when using the projection product. Figure 2 also shows that a similar effect occurs in all five of the language models studied in this research, indicating that this method allows for the measurement of human-interpretable semantics and bias in highly contextual and anisotropic embedding spaces.

\subsection{Measuring Differential Bias Based on Valence}

As shown in Table \ref{tab:valence_biases}, the evidence suggests that language models encode consistent valence biases based on gender identity, sexual orientation, and social class signals in an intersectional context. A statistically significant positive effect size is obtained for the heterosexual vs. homosexual and cisgender vs. transgender test for all five of the models studied in this research. For ALBERT and RoBERTa, effect sizes are large ($d = 1.34$ and $d = 1.22$) for the gender identity test; medium effect sizes are obtained for GPT-Neo ($d = 0.64$) and T5 ($d = 0.61$) for the sexual orientation test. Statistically significant valence bias effect sizes are also obtained for four language models for the affluent vs. destitute test. Bias effect sizes are medium ($d > 0.5$) or large ($d > 0.8$) in three of the five models. The large effect size for social class speaks to the presence of biases related to social class in language models, a relatively unexplored bias type in AI except for the work of \citet{doi:10.1177/0003122419877135} analyzing the meaning of class in static word embeddings. Figure \ref{fig:avg1} visualizes the difference in the mean projection onto the valence dimension for each of the $12$ biases studied.

Another noteworthy result is that three of five language models (ALBERT, GPT-Neo, and RoBERTa) differentially associate men with pleasantness over women. While effect sizes are small, this deviates from psychological research suggesting that women are evaluated as more pleasant than men. For example, while men are often associated with aggression and violence, women are associated with more communal attributes such as warmth. This is known as the "women-are-wonderful-effect" \cite{doi:10.1111/j.1471-6402.1991.tb00792.x, articlewom, doi:10.1080/14792779543000002, EAGLY1994113}. It is possible that women are portrayed negatively in the training corpora of these language models, causing men to be more differentially pleasant. This possibility is supported by the recent research of \citet{birhane2021multimodal}, who find that corpora used for training language-and-image models contain misogynistic and toxically stereotypical depictions of women. The association of women with pleasantness is, however, observed in T5.

Results across five language models suggest the utility of the method proposed in this research for capturing widespread societal biases in contextualized word embeddings. Bias effect sizes are stereotype-congruent in at least $9$ of $12$ tests for three of the five models assessed, and in every model at least half of the bias tests yield positive effect sizes. Moreover, the results presented here further affirm the findings of \citet{nadeem2020stereoset}, who find that larger language models are both better at language modeling and more biased based on a downstream evaluation of bias. The present research observes that GPT-Neo, the largest of the language models studied herein and previously observed to outperform other language models on both intrinsic and downstream evaluations of semantic quality \cite{wolfe2022vast,gao2020pile}, has a statistically significant bias effect size of at least $0.50$ for $6$ of the $12$ bias tests, the most of any of the models studied herein.

\subsection{Identifying the Strongest Affective Biases in a Language Model}

\begin{figure*}[!htbp]
\begin{tikzpicture}
\begin{axis} [
    height=4cm,
    width=15cm,
    ybar = .05cm,
    bar width = 8pt,
    ymin = 0, 
    ymax = 100,
    ylabel= \% of Contexts,
    ylabel near ticks,
    xtick = {1,2,3,4,5,6,7,8,9,10},
    xtick style={draw=none},
    ytick pos = left,
    xticklabels = {Heterosexual, Cisgender, Christian, Black, Female, Male, White, Muslim, Transgender, Homosexual},
    xticklabel style={rotate=20,anchor=east},
    x label style={at={(axis description cs:0.5,-0.1)},anchor=north},
    legend style={at={(0.87,1)},anchor=south west},
    title=Occurrences in 10\% Most Pleasant and Unpleasant GPT-Neo Contexts,
    xlabel= {Social Category},
    enlarge x limits={abs=1cm},
]

\addplot [pattern=grid,pattern color = blue] coordinates {(1,93.75) (2,70.05) (3,65.10) (4,63.54) (5,52.86) (6,47.14) (7,36.46) (8,34.90) (9,29.95) (10,6.25)};

\addplot [pattern=dots,pattern color = orange]  coordinates {(1,0.26) (2,6.77) (3,49.22) (4,45.05) (5,48.44) (6,51.56) (7,54.95) (8,50.78) (9,93.23) (10,99.74)};

\legend {Most Pleasant, Most Unpleasant};
\end{axis}
\end{tikzpicture}
\caption{More than 93\% of the most pleasant contextualized representations of the word "person" in the top layer of GPT-Neo include the word "heterosexual," and more than 70\% include the word "cisgender." On the other hand, more than 99\% of the most unpleasant phrases in the top layer of GPT-Neo include the word "homosexual," and more than 93\% include the word "transgender," reflecting biases based on sexual orientation and gender identity.}
\label{fig:top_10_pct}
\end{figure*}
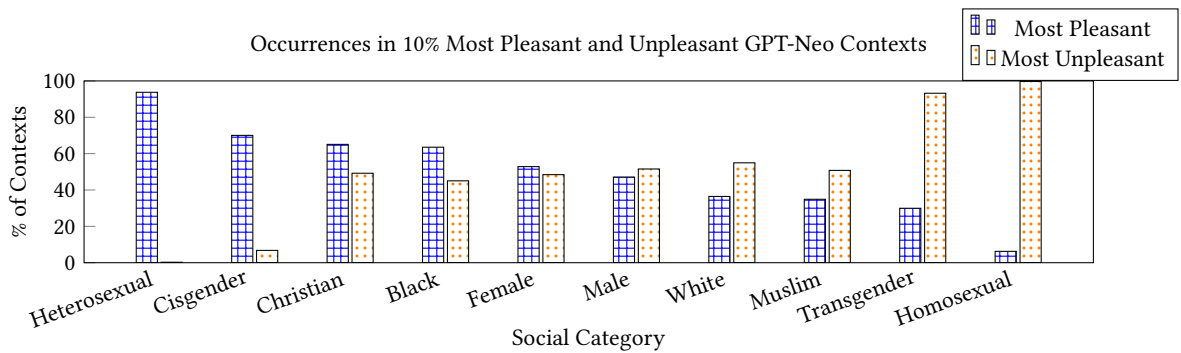

As shown in Figure \ref{fig:top_10_pct}, retrieving the top 10\% of the most pleasant contexts shows that heterosexuality and cisgender identity are over-represented in the most positively valenced phrases in GPT-Neo, with more than 93\% of the most pleasant phrases containing the word "heterosexual," and more than 70\% of the most pleasant phrases containing the word "cisgender." The word "Christian" is also positively valenced, with more than 65\% of the most pleasant phrases containing the word. On the other hand, the word "homosexual" occurs in the most positively valenced phrases less than 7\% of the time, and retrieving the top 10\% of the most unpleasant contexts shows that homosexuality and transgender identity are among the most negatively valenced words assessed, with more than 99\% of the most negative phrases containing the word "homosexual," and more than 93\% of the most negative phrases containing the word "transgender." None of the eight other words assessed occurs in more than 55\% of the most negative phrases. The word "heterosexual" occurs less than 1\% of the time in the most negatively valenced phrases. The word "Muslim" occurs more frequently in the most negatively valenced phrases than it does in the most positively valenced phrases, as does the word "white." The words "male" and "female" occur roughly equally in the most positively and negatively valenced phrases.

Both figures~\ref{fig:avg1} and \ref{fig:top_10_pct} show that a "white" person is slightly more negatively valenced than a "black" person in GPT-Neo ($d = -0.12$ in Table~\ref{tab:valence_biases}). In representational models, such as multi-modal vision-language models, the default unmarked person in English is associated with "white" \cite{wolfe2022american}, as a result the noun the person does not typically get marked with the identity descriptor of "white" \cite{wolfe2022markedness}. The effect of markedness for a "black" person might potentially be causing the stereotype incongruent result.

The results of this method, which does not require the definition of binary groups for differential measurement, are mostly consistent with the results obtained from the differential statistical test introduced in the second experiment. This suggests the utility of the projection method for measuring biases in contextualized word embeddings even when an opposing category does not exist such that a differential bias test can be performed.

\section{Discussion}

The contributions of the present research are threefold: a method for measuring semantics in contextual and anisotropic embedding spaces; a novel and generalizable differential bias measurement which takes into account the contextualization property of all language models, and which returns an effect size indicating magnitude and a $p$-value measuring statistical significance; and a means for quantifying biases in contextualized word embeddings in an intersectional setting. By analyzing sexual orientation, social class, and gender bias without having to use gender binary, our approach is more inclusive compared to various previous analyses.  

The findings of this work indicate that the biases demonstrating low regard based on sexual orientation observed by \citet{sheng2019woman} in the text output of language models can be traced back to the contextualized embedding space, where the occurrence of the words "homosexual" and "transgender" lead to greater association with unpleasantness and negative attitudes. Future work might use the method put forth in this research to further examine the link between bias in contextualized word embeddings and the propagation of that bias to model objectives such as language generation or other downstream NLP tasks such as sentiment analysis, machine translation \cite{ghosh2023chatgpt}, or consequential decision making.

Many of the results reported in this work suggest that biases of contextualization have consistent indirect impacts on the representation of societally disadvantaged people. For example, the results indicate that biases related to education and social class exist in many language models. These categories speak directly to the opportunities and outcomes afforded over which an individual typically has little control.

Moreover, many of the biases observed in this work are likely to interact with and intensify other biases. For example, a bias based on sex (male vs. female) is observed in most language models, such that men are more associated with pleasantness than women. However, biases based on weight, and age are also observed, such that the word person is more pleasant when it occurs with "thin," and "young." While such biases may affect any context in which they are observed, they are likely to have greater impact on the representation of women in language models, as women are more likely at a societal level to be described with regard to their physical appearances, and biases related to age are often directed more strongly toward women, and at younger ages than men \cite{harris1994growing}. The consequence of the contextualization effect observed in this research is that representations of people more likely to be described in a biased manner will become even more negatively valenced in the model than the categorical biases indicate when considered individually.

The methods described in this research have ramifications not only for studies of bias in AI, but also for the social sciences, as social scientists may use the computational approaches described in this research to quantify properties of human language and culture, without the problem of meaning being collapsed into a single vector representation, as occurs in static word embeddings. While norms and biases based on valence are studied in this work to ground a new method in prior psychological research, a maximum margin subspace could be learned to represent many other semantic properties; for example, future research might learn a political spectrum subspace to study biases beyond those observable based on valence.

Finally, while this work assesses in-context biases, it evaluates their impact in individual or two differential categories. However, the method can be trivially extended such that intersectional identities can be assessed by observing biases based on word bigrams, trigrams, or longer descriptive sequences. This is facilitated by using the contextualized representation of the word "person" as the target embedding for all bias measurements, rather than attempting to directly measure the embedded representations of bias-inducing words or categories.
\subsection{Limitations and Future Work}

The results reported for \hyperref[sec:eval-learn]{experiment 5.2} are obtained by generating combinations of categories representing $12$ social biases. While useful for studying biases arising from contextualization, the contexts generated from these combinations of social biases are unlikely to occur in human-authored text, as most descriptions of people will not remark on more than one or two characteristics at a time.

Future work might explore the use of this method in more natural contexts, perhaps similar to the approach used by \citet{wolfe2021low}, who study racial and gender biases related to names by interchanging names in otherwise identical contexts derived from human-authored sources. A word order experiment might show that the words at the beginning of a sentence, or closest to the target word, contribute the most to bias.

\section{Conclusion}

This research introduces a novel and effective machine learning approach to measuring valence associations in contextualized word embeddings. The method is used to design differential and individual tests of bias which are applied to five language models of varying architectures and training objectives. Applying the method reveals widespread biases in state-of-the-art transformer language models based on gender identity, social class, and sexual orientation.

\begin{acks}
This work is supported by the U.S. National Institute of Standards and Technology (NIST) Grant 60NANB20D212T. Any opinions, findings, and conclusions or recommendations expressed in this material are those of the author and do not necessarily reflect those of NIST.
\end{acks}
 
\bibliographystyle{ACM-Reference-Format}
\bibliography{sample-base}


\begin{thebibliography}{83}


\ifx \showCODEN    \undefined \def \showCODEN     #1{\unskip}     \fi
\ifx \showDOI      \undefined \def \showDOI       #1{#1}\fi
\ifx \showISBNx    \undefined \def \showISBNx     #1{\unskip}     \fi
\ifx \showISBNxiii \undefined \def \showISBNxiii  #1{\unskip}     \fi
\ifx \showISSN     \undefined \def \showISSN      #1{\unskip}     \fi
\ifx \showLCCN     \undefined \def \showLCCN      #1{\unskip}     \fi
\ifx \shownote     \undefined \def \shownote      #1{#1}          \fi
\ifx \showarticletitle \undefined \def \showarticletitle #1{#1}   \fi
\ifx \showURL      \undefined \def \showURL       {\relax}        \fi
\providecommand\bibfield[2]{#2}
\providecommand\bibinfo[2]{#2}
\providecommand\natexlab[1]{#1}
\providecommand\showeprint[2][]{arXiv:#2}

\bibitem[\protect\citeauthoryear{Basta, Costa{-}juss{\`{a}}, and Casas}{Basta
  et~al\mbox{.}}{2019}]%
        {DBLP:journals/corr/abs-1904-08783}
\bibfield{author}{\bibinfo{person}{Christine Basta},
  \bibinfo{person}{Marta~Ruiz Costa{-}juss{\`{a}}}, {and} \bibinfo{person}{Noe
  Casas}.} \bibinfo{year}{2019}\natexlab{}.
\newblock \showarticletitle{Evaluating the Underlying Gender Bias in
  Contextualized Word Embeddings}.
\newblock \bibinfo{journal}{\emph{CoRR}}  \bibinfo{volume}{abs/1904.08783}
  (\bibinfo{year}{2019}).
\newblock
\showeprint[arxiv]{1904.08783}
\urldef\tempurl%
\url{http://arxiv.org/abs/1904.08783}
\showURL{%
\tempurl}


\bibitem[\protect\citeauthoryear{Bellezza, Greenwald, and Banaji}{Bellezza
  et~al\mbox{.}}{1986}]%
        {RefWorks:doc:613a1f068f08743fdb7624b5}
\bibfield{author}{\bibinfo{person}{Francis~S. Bellezza},
  \bibinfo{person}{Anthony~G. Greenwald}, {and} \bibinfo{person}{Mahzarin~R.
  Banaji}.} \bibinfo{year}{1986}\natexlab{}.
\newblock \showarticletitle{Words high and low in pleasantness as rated by male
  and female college students}.
\newblock \bibinfo{journal}{\emph{Behavior Research Methods, Instruments \&
  Computers}} \bibinfo{volume}{18}, \bibinfo{number}{3} (\bibinfo{year}{1986}),
  \bibinfo{pages}{299--303}.
\newblock
\showISBNx{0743-3808(PRINT)}
\urldef\tempurl%
\url{https://doi.org/10.3758/BF03204403}
\showDOI{\tempurl}
\newblock
\shownote{ID: 1988-03937-001.}


\bibitem[\protect\citeauthoryear{Bengio, Ducharme, Vincent, and Janvin}{Bengio
  et~al\mbox{.}}{2003}]%
        {bengio2003neural}
\bibfield{author}{\bibinfo{person}{Yoshua Bengio}, \bibinfo{person}{R{\'e}jean
  Ducharme}, \bibinfo{person}{Pascal Vincent}, {and} \bibinfo{person}{Christian
  Janvin}.} \bibinfo{year}{2003}\natexlab{}.
\newblock \showarticletitle{A neural probabilistic language model}.
\newblock \bibinfo{journal}{\emph{The journal of machine learning research}}
  \bibinfo{volume}{3} (\bibinfo{year}{2003}), \bibinfo{pages}{1137--1155}.
\newblock


\bibitem[\protect\citeauthoryear{Birhane, Prabhu, and Kahembwe}{Birhane
  et~al\mbox{.}}{2021}]%
        {birhane2021multimodal}
\bibfield{author}{\bibinfo{person}{Abeba Birhane}, \bibinfo{person}{Vinay~Uday
  Prabhu}, {and} \bibinfo{person}{Emmanuel Kahembwe}.}
  \bibinfo{year}{2021}\natexlab{}.
\newblock \showarticletitle{Multimodal datasets: misogyny, pornography, and
  malignant stereotypes}.
\newblock \bibinfo{journal}{\emph{arXiv preprint arXiv:2110.01963}}
  (\bibinfo{year}{2021}).
\newblock


\bibitem[\protect\citeauthoryear{Black, Gao, Wang, Leahy, and Biderman}{Black
  et~al\mbox{.}}{2021}]%
        {gpt-neo}
\bibfield{author}{\bibinfo{person}{Sid Black}, \bibinfo{person}{Leo Gao},
  \bibinfo{person}{Phil Wang}, \bibinfo{person}{Connor Leahy}, {and}
  \bibinfo{person}{Stella Biderman}.} \bibinfo{year}{2021}\natexlab{}.
\newblock \bibinfo{booktitle}{\emph{{GPT-Neo}: Large Scale Autoregressive
  Language Modeling with Mesh-Tensorflow}}.
\newblock
\urldef\tempurl%
\url{https://doi.org/10.5281/zenodo.5297715}
\showDOI{\tempurl}


\bibitem[\protect\citeauthoryear{Bojanowski, Grave, Joulin, and
  Mikolov}{Bojanowski et~al\mbox{.}}{2017}]%
        {bojanowski2017enriching}
\bibfield{author}{\bibinfo{person}{Piotr Bojanowski}, \bibinfo{person}{Edouard
  Grave}, \bibinfo{person}{Armand Joulin}, {and} \bibinfo{person}{Tomas
  Mikolov}.} \bibinfo{year}{2017}\natexlab{}.
\newblock \bibinfo{title}{Enriching Word Vectors with Subword Information}.
\newblock
\newblock
\showeprint[arxiv]{1607.04606}~[cs.CL]


\bibitem[\protect\citeauthoryear{Bolukbasi, Chang, Zou, Saligrama, and
  Kalai}{Bolukbasi et~al\mbox{.}}{2016}]%
        {10.5555/3157382.3157584}
\bibfield{author}{\bibinfo{person}{Tolga Bolukbasi}, \bibinfo{person}{Kai-Wei
  Chang}, \bibinfo{person}{James Zou}, \bibinfo{person}{Venkatesh Saligrama},
  {and} \bibinfo{person}{Adam Kalai}.} \bibinfo{year}{2016}\natexlab{}.
\newblock \showarticletitle{Man is to Computer Programmer as Woman is to
  Homemaker? Debiasing Word Embeddings}. In
  \bibinfo{booktitle}{\emph{Proceedings of the 30th International Conference on
  Neural Information Processing Systems}} (Barcelona, Spain)
  \emph{(\bibinfo{series}{NIPS'16})}. \bibinfo{publisher}{Curran Associates
  Inc.}, \bibinfo{address}{Red Hook, NY, USA}, \bibinfo{pages}{4356–4364}.
\newblock
\showISBNx{9781510838819}


\bibitem[\protect\citeauthoryear{Bradley and Lang}{Bradley and Lang}{1999}]%
        {Bradley1999AffectiveNF}
\bibfield{author}{\bibinfo{person}{Margaret~M. Bradley} {and}
  \bibinfo{person}{Peter~J. Lang}.} \bibinfo{year}{1999}\natexlab{}.
\newblock \showarticletitle{Affective Norms for English Words (ANEW):
  Instruction Manual and Affective Ratings}.
\newblock


\bibitem[\protect\citeauthoryear{Brown, Mann, Ryder, Subbiah, Kaplan, Dhariwal,
  Neelakantan, Shyam, Sastry, Askell, Agarwal, Herbert-Voss, Krueger, Henighan,
  Child, Ramesh, Ziegler, Wu, Winter, Hesse, Chen, Sigler, Litwin, Gray, Chess,
  Clark, Berner, McCandlish, Radford, Sutskever, and Amodei}{Brown
  et~al\mbox{.}}{2020}]%
        {NEURIPS2020_1457c0d6}
\bibfield{author}{\bibinfo{person}{Tom Brown}, \bibinfo{person}{Benjamin Mann},
  \bibinfo{person}{Nick Ryder}, \bibinfo{person}{Melanie Subbiah},
  \bibinfo{person}{Jared~D Kaplan}, \bibinfo{person}{Prafulla Dhariwal},
  \bibinfo{person}{Arvind Neelakantan}, \bibinfo{person}{Pranav Shyam},
  \bibinfo{person}{Girish Sastry}, \bibinfo{person}{Amanda Askell},
  \bibinfo{person}{Sandhini Agarwal}, \bibinfo{person}{Ariel Herbert-Voss},
  \bibinfo{person}{Gretchen Krueger}, \bibinfo{person}{Tom Henighan},
  \bibinfo{person}{Rewon Child}, \bibinfo{person}{Aditya Ramesh},
  \bibinfo{person}{Daniel Ziegler}, \bibinfo{person}{Jeffrey Wu},
  \bibinfo{person}{Clemens Winter}, \bibinfo{person}{Chris Hesse},
  \bibinfo{person}{Mark Chen}, \bibinfo{person}{Eric Sigler},
  \bibinfo{person}{Mateusz Litwin}, \bibinfo{person}{Scott Gray},
  \bibinfo{person}{Benjamin Chess}, \bibinfo{person}{Jack Clark},
  \bibinfo{person}{Christopher Berner}, \bibinfo{person}{Sam McCandlish},
  \bibinfo{person}{Alec Radford}, \bibinfo{person}{Ilya Sutskever}, {and}
  \bibinfo{person}{Dario Amodei}.} \bibinfo{year}{2020}\natexlab{}.
\newblock \showarticletitle{Language Models are Few-Shot Learners}. In
  \bibinfo{booktitle}{\emph{Advances in Neural Information Processing
  Systems}}, \bibfield{editor}{\bibinfo{person}{H.~Larochelle},
  \bibinfo{person}{M.~Ranzato}, \bibinfo{person}{R.~Hadsell},
  \bibinfo{person}{M.~F. Balcan}, {and} \bibinfo{person}{H.~Lin}} (Eds.),
  Vol.~\bibinfo{volume}{33}. \bibinfo{publisher}{Curran Associates, Inc.},
  \bibinfo{pages}{1877--1901}.
\newblock
\urldef\tempurl%
\url{https://proceedings.neurips.cc/paper/2020/file/1457c0d6bfcb4967418bfb8ac142f64a-Paper.pdf}
\showURL{%
\tempurl}


\bibitem[\protect\citeauthoryear{Buck, Heafield, and Van~Ooyen}{Buck
  et~al\mbox{.}}{2014}]%
        {buck2014n}
\bibfield{author}{\bibinfo{person}{Christian Buck}, \bibinfo{person}{Kenneth
  Heafield}, {and} \bibinfo{person}{Bas Van~Ooyen}.}
  \bibinfo{year}{2014}\natexlab{}.
\newblock \showarticletitle{N-gram counts and language models from the common
  crawl}. In \bibinfo{booktitle}{\emph{Proceedings of the Ninth International
  Conference on Language Resources and Evaluation (LREC'14)}}.
  \bibinfo{pages}{3579--3584}.
\newblock


\bibitem[\protect\citeauthoryear{Caliskan}{Caliskan}{2021}]%
        {caliskan2021detecting}
\bibfield{author}{\bibinfo{person}{Aylin Caliskan}.}
  \bibinfo{year}{2021}\natexlab{}.
\newblock \showarticletitle{Detecting and mitigating bias in natural language
  processing}.
\newblock \bibinfo{journal}{\emph{Brookings Institution}}
  (\bibinfo{year}{2021}).
\newblock


\bibitem[\protect\citeauthoryear{Caliskan, Bryson, and Narayanan}{Caliskan
  et~al\mbox{.}}{2017}]%
        {Caliskan_2017}
\bibfield{author}{\bibinfo{person}{Aylin Caliskan}, \bibinfo{person}{Joanna~J.
  Bryson}, {and} \bibinfo{person}{Arvind Narayanan}.}
  \bibinfo{year}{2017}\natexlab{}.
\newblock \showarticletitle{Semantics derived automatically from language
  corpora contain human-like biases}.
\newblock \bibinfo{journal}{\emph{Science}} \bibinfo{volume}{356},
  \bibinfo{number}{6334} (\bibinfo{date}{Apr} \bibinfo{year}{2017}),
  \bibinfo{pages}{183–186}.
\newblock
\showISSN{1095-9203}
\urldef\tempurl%
\url{https://doi.org/10.1126/science.aal4230}
\showDOI{\tempurl}


\bibitem[\protect\citeauthoryear{Caliskan and Lewis}{Caliskan and
  Lewis}{[n.d.]}]%
        {caliskan2020social}
\bibfield{author}{\bibinfo{person}{Aylin Caliskan} {and} \bibinfo{person}{Molly
  Lewis}.} \bibinfo{year}{[n.d.]}\natexlab{}.
\newblock \showarticletitle{Social biases in word embeddings and their relation
  to human cognition}.
\newblock  (\bibinfo{year}{[n.\,d.]}).
\newblock


\bibitem[\protect\citeauthoryear{Callan, Hoy, Yoo, and Zhao}{Callan
  et~al\mbox{.}}{2009}]%
        {callan2009clueweb09}
\bibfield{author}{\bibinfo{person}{Jamie Callan}, \bibinfo{person}{Mark Hoy},
  \bibinfo{person}{Changkuk Yoo}, {and} \bibinfo{person}{Le Zhao}.}
  \bibinfo{year}{2009}\natexlab{}.
\newblock \bibinfo{title}{Clueweb09 data set}.
\newblock
\newblock


\bibitem[\protect\citeauthoryear{Charlesworth, Caliskan, and
  Banaji}{Charlesworth et~al\mbox{.}}{2022}]%
        {charlesworth2022historical}
\bibfield{author}{\bibinfo{person}{Tessa~ES Charlesworth},
  \bibinfo{person}{Aylin Caliskan}, {and} \bibinfo{person}{Mahzarin~R Banaji}.}
  \bibinfo{year}{2022}\natexlab{}.
\newblock \showarticletitle{Historical representations of social groups across
  200 years of word embeddings from Google Books}.
\newblock \bibinfo{journal}{\emph{Proceedings of the National Academy of
  Sciences}} \bibinfo{volume}{119}, \bibinfo{number}{28}
  (\bibinfo{year}{2022}), \bibinfo{pages}{e2121798119}.
\newblock


\bibitem[\protect\citeauthoryear{Cheng, Durmus, and Jurafsky}{Cheng
  et~al\mbox{.}}{2023}]%
        {cheng2023marked}
\bibfield{author}{\bibinfo{person}{Myra Cheng}, \bibinfo{person}{Esin Durmus},
  {and} \bibinfo{person}{Dan Jurafsky}.} \bibinfo{year}{2023}\natexlab{}.
\newblock \bibinfo{title}{Marked Personas: Using Natural Language Prompts to
  Measure Stereotypes in Language Models}.
\newblock
\newblock
\showeprint[arxiv]{2305.18189}~[cs.CL]


\bibitem[\protect\citeauthoryear{Devlin, Chang, Lee, and Toutanova}{Devlin
  et~al\mbox{.}}{2019}]%
        {devlin-etal-2019-bert}
\bibfield{author}{\bibinfo{person}{Jacob Devlin}, \bibinfo{person}{Ming-Wei
  Chang}, \bibinfo{person}{Kenton Lee}, {and} \bibinfo{person}{Kristina
  Toutanova}.} \bibinfo{year}{2019}\natexlab{}.
\newblock \showarticletitle{{BERT}: Pre-training of Deep Bidirectional
  Transformers for Language Understanding}. In
  \bibinfo{booktitle}{\emph{Proceedings of the 2019 Conference of the North
  {A}merican Chapter of the Association for Computational Linguistics: Human
  Language Technologies, Volume 1 (Long and Short Papers)}}.
  \bibinfo{publisher}{Association for Computational Linguistics},
  \bibinfo{address}{Minneapolis, Minnesota}, \bibinfo{pages}{4171--4186}.
\newblock
\urldef\tempurl%
\url{https://doi.org/10.18653/v1/N19-1423}
\showDOI{\tempurl}


\bibitem[\protect\citeauthoryear{Eagly and Mladinic}{Eagly and
  Mladinic}{1989}]%
        {articlewom}
\bibfield{author}{\bibinfo{person}{Alice Eagly} {and} \bibinfo{person}{Antonio
  Mladinic}.} \bibinfo{year}{1989}\natexlab{}.
\newblock \showarticletitle{Gender Stereotypes and Attitudes Toward Women and
  Men}.
\newblock \bibinfo{journal}{\emph{Personality and Social Psychology Bulletin}}
  \bibinfo{volume}{15} (\bibinfo{date}{12} \bibinfo{year}{1989}),
  \bibinfo{pages}{543--558}.
\newblock
\urldef\tempurl%
\url{https://doi.org/10.1177/0146167289154008}
\showDOI{\tempurl}


\bibitem[\protect\citeauthoryear{Eagly and Mladinic}{Eagly and
  Mladinic}{1994}]%
        {doi:10.1080/14792779543000002}
\bibfield{author}{\bibinfo{person}{Alice~H. Eagly} {and}
  \bibinfo{person}{Antonio Mladinic}.} \bibinfo{year}{1994}\natexlab{}.
\newblock \showarticletitle{Are People Prejudiced Against Women? Some Answers
  From Research on Attitudes, Gender Stereotypes, and Judgments of Competence}.
\newblock \bibinfo{journal}{\emph{European Review of Social Psychology}}
  \bibinfo{volume}{5}, \bibinfo{number}{1} (\bibinfo{year}{1994}),
  \bibinfo{pages}{1--35}.
\newblock
\urldef\tempurl%
\url{https://doi.org/10.1080/14792779543000002}
\showDOI{\tempurl}
\showeprint{https://doi.org/10.1080/14792779543000002}


\bibitem[\protect\citeauthoryear{Eagly, Mladinic, and Otto}{Eagly
  et~al\mbox{.}}{1991}]%
        {doi:10.1111/j.1471-6402.1991.tb00792.x}
\bibfield{author}{\bibinfo{person}{Alice~H. Eagly}, \bibinfo{person}{Antonio
  Mladinic}, {and} \bibinfo{person}{Stacey Otto}.}
  \bibinfo{year}{1991}\natexlab{}.
\newblock \showarticletitle{Are Women Evaluated More Favorably Than Men?: An
  Analysis of Attitudes, Beliefs, and Emotions}.
\newblock \bibinfo{journal}{\emph{Psychology of Women Quarterly}}
  \bibinfo{volume}{15}, \bibinfo{number}{2} (\bibinfo{year}{1991}),
  \bibinfo{pages}{203--216}.
\newblock
\urldef\tempurl%
\url{https://doi.org/10.1111/j.1471-6402.1991.tb00792.x}
\showDOI{\tempurl}
\showeprint{https://doi.org/10.1111/j.1471-6402.1991.tb00792.x}


\bibitem[\protect\citeauthoryear{Eagly, Mladinic, and Otto}{Eagly
  et~al\mbox{.}}{1994}]%
        {EAGLY1994113}
\bibfield{author}{\bibinfo{person}{Alice~H. Eagly}, \bibinfo{person}{Antonio
  Mladinic}, {and} \bibinfo{person}{Stacey Otto}.}
  \bibinfo{year}{1994}\natexlab{}.
\newblock \showarticletitle{Cognitive and Affective Bases of Attitudes toward
  Social Groups and Social Policies}.
\newblock \bibinfo{journal}{\emph{Journal of Experimental Social Psychology}}
  \bibinfo{volume}{30}, \bibinfo{number}{2} (\bibinfo{year}{1994}),
  \bibinfo{pages}{113--137}.
\newblock
\showISSN{0022-1031}
\urldef\tempurl%
\url{https://doi.org/10.1006/jesp.1994.1006}
\showDOI{\tempurl}


\bibitem[\protect\citeauthoryear{Ethayarajh}{Ethayarajh}{2019}]%
        {ethayarajh-2019-contextual}
\bibfield{author}{\bibinfo{person}{Kawin Ethayarajh}.}
  \bibinfo{year}{2019}\natexlab{}.
\newblock \showarticletitle{How Contextual are Contextualized Word
  Representations? {C}omparing the Geometry of {BERT}, {ELM}o, and {GPT}-2
  Embeddings}. In \bibinfo{booktitle}{\emph{Proceedings of the 2019 Conference
  on Empirical Methods in Natural Language Processing and the 9th International
  Joint Conference on Natural Language Processing (EMNLP-IJCNLP)}}.
  \bibinfo{publisher}{Association for Computational Linguistics},
  \bibinfo{address}{Hong Kong, China}, \bibinfo{pages}{55--65}.
\newblock
\urldef\tempurl%
\url{https://doi.org/10.18653/v1/D19-1006}
\showDOI{\tempurl}


\bibitem[\protect\citeauthoryear{Gao, Biderman, Black, Golding, Hoppe, Foster,
  Phang, He, Thite, Nabeshima, et~al\mbox{.}}{Gao et~al\mbox{.}}{2020}]%
        {gao2020pile}
\bibfield{author}{\bibinfo{person}{Leo Gao}, \bibinfo{person}{Stella Biderman},
  \bibinfo{person}{Sid Black}, \bibinfo{person}{Laurence Golding},
  \bibinfo{person}{Travis Hoppe}, \bibinfo{person}{Charles Foster},
  \bibinfo{person}{Jason Phang}, \bibinfo{person}{Horace He},
  \bibinfo{person}{Anish Thite}, \bibinfo{person}{Noa Nabeshima},
  {et~al\mbox{.}}} \bibinfo{year}{2020}\natexlab{}.
\newblock \showarticletitle{The Pile: An 800GB Dataset of Diverse Text for
  Language Modeling}.
\newblock \bibinfo{journal}{\emph{arXiv preprint arXiv:2101.00027}}
  (\bibinfo{year}{2020}).
\newblock


\bibitem[\protect\citeauthoryear{Garg, Schiebinger, Jurafsky, and Zou}{Garg
  et~al\mbox{.}}{2018}]%
        {GargE3635}
\bibfield{author}{\bibinfo{person}{Nikhil Garg}, \bibinfo{person}{Londa
  Schiebinger}, \bibinfo{person}{Dan Jurafsky}, {and} \bibinfo{person}{James
  Zou}.} \bibinfo{year}{2018}\natexlab{}.
\newblock \showarticletitle{Word embeddings quantify 100 years of gender and
  ethnic stereotypes}.
\newblock \bibinfo{journal}{\emph{Proceedings of the National Academy of
  Sciences}} \bibinfo{volume}{115}, \bibinfo{number}{16}
  (\bibinfo{year}{2018}), \bibinfo{pages}{E3635--E3644}.
\newblock
\showISSN{0027-8424}
\urldef\tempurl%
\url{https://doi.org/10.1073/pnas.1720347115}
\showDOI{\tempurl}
\showeprint{https://www.pnas.org/content/115/16/E3635.full.pdf}


\bibitem[\protect\citeauthoryear{Ghosh and Caliskan}{Ghosh and
  Caliskan}{2023}]%
        {ghosh2023chatgpt}
\bibfield{author}{\bibinfo{person}{Sourojit Ghosh} {and} \bibinfo{person}{Aylin
  Caliskan}.} \bibinfo{year}{2023}\natexlab{}.
\newblock \showarticletitle{ChatGPT Perpetuates Gender Bias in Machine
  Translation and Ignores Non-Gendered Pronouns: Findings across Bengali and
  Five other Low-Resource Languages}.
\newblock \bibinfo{journal}{\emph{arXiv preprint arXiv:2305.10510}}
  (\bibinfo{year}{2023}).
\newblock


\bibitem[\protect\citeauthoryear{Gokaslan and Cohen}{Gokaslan and
  Cohen}{2019}]%
        {Gokaslan2019OpenWeb}
\bibfield{author}{\bibinfo{person}{Aaron Gokaslan} {and} \bibinfo{person}{Vanya
  Cohen}.} \bibinfo{year}{2019}\natexlab{}.
\newblock \bibinfo{title}{OpenWebText Corpus}.
\newblock
  \bibinfo{howpublished}{\url{http://Skylion007.github.io/OpenWebTextCorpus}}.
\newblock


\bibitem[\protect\citeauthoryear{Gonen and Goldberg}{Gonen and
  Goldberg}{2019}]%
        {GONEN19}
\bibfield{author}{\bibinfo{person}{Hila Gonen} {and} \bibinfo{person}{Yoav
  Goldberg}.} \bibinfo{year}{2019}\natexlab{}.
\newblock \showarticletitle{Lipstick on a Pig: Debiasing Methods Cover up
  Systematic Gender Biases in Word Embeddings But do not Remove Them}. In
  \bibinfo{booktitle}{\emph{Proceedings of NAACL-HLT}}.
\newblock


\bibitem[\protect\citeauthoryear{Greenwald, McGhee, and Schwartz}{Greenwald
  et~al\mbox{.}}{1998}]%
        {greenwald1998measuring}
\bibfield{author}{\bibinfo{person}{Anthony~G Greenwald},
  \bibinfo{person}{Debbie~E McGhee}, {and} \bibinfo{person}{Jordan~LK
  Schwartz}.} \bibinfo{year}{1998}\natexlab{}.
\newblock \showarticletitle{Measuring individual differences in implicit
  cognition: the implicit association test.}
\newblock \bibinfo{journal}{\emph{Journal of personality and social
  psychology}} \bibinfo{volume}{74}, \bibinfo{number}{6}
  (\bibinfo{year}{1998}), \bibinfo{pages}{1464}.
\newblock


\bibitem[\protect\citeauthoryear{Guo and Caliskan}{Guo and Caliskan}{2021}]%
        {10.1145/3461702.3462536}
\bibfield{author}{\bibinfo{person}{Wei Guo} {and} \bibinfo{person}{Aylin
  Caliskan}.} \bibinfo{year}{2021}\natexlab{}.
\newblock \showarticletitle{Detecting Emergent Intersectional Biases:
  Contextualized Word Embeddings Contain a Distribution of Human-like Biases}.
  In \bibinfo{booktitle}{\emph{Proceedings of the 2021 AAAI/ACM Conference on
  AI, Ethics, and Society}} (Virtual Event, USA) \emph{(\bibinfo{series}{AIES
  '21})}. \bibinfo{publisher}{Association for Computing Machinery},
  \bibinfo{address}{New York, NY, USA}, \bibinfo{pages}{122–133}.
\newblock
\showISBNx{9781450384735}
\urldef\tempurl%
\url{https://doi.org/10.1145/3461702.3462536}
\showDOI{\tempurl}


\bibitem[\protect\citeauthoryear{Hamilton, Leskovec, and Jurafsky}{Hamilton
  et~al\mbox{.}}{2016}]%
        {hamilton-etal-2016-diachronic}
\bibfield{author}{\bibinfo{person}{William~L. Hamilton}, \bibinfo{person}{Jure
  Leskovec}, {and} \bibinfo{person}{Dan Jurafsky}.}
  \bibinfo{year}{2016}\natexlab{}.
\newblock \showarticletitle{Diachronic Word Embeddings Reveal Statistical Laws
  of Semantic Change}. In \bibinfo{booktitle}{\emph{Proceedings of the 54th
  Annual Meeting of the Association for Computational Linguistics (Volume 1:
  Long Papers)}}. \bibinfo{publisher}{Association for Computational
  Linguistics}, \bibinfo{address}{Berlin, Germany},
  \bibinfo{pages}{1489--1501}.
\newblock
\urldef\tempurl%
\url{https://doi.org/10.18653/v1/P16-1141}
\showDOI{\tempurl}


\bibitem[\protect\citeauthoryear{Harris}{Harris}{1994}]%
        {harris1994growing}
\bibfield{author}{\bibinfo{person}{Mary~B Harris}.}
  \bibinfo{year}{1994}\natexlab{}.
\newblock \showarticletitle{Growing old gracefully: Age concealment and
  gender}.
\newblock \bibinfo{journal}{\emph{Journal of Gerontology}}
  \bibinfo{volume}{49}, \bibinfo{number}{4} (\bibinfo{year}{1994}),
  \bibinfo{pages}{P149--P158}.
\newblock


\bibitem[\protect\citeauthoryear{Hogg and Abrams}{Hogg and Abrams}{2007}]%
        {kar23659}
\bibfield{author}{\bibinfo{person}{Michael~A. Hogg} {and}
  \bibinfo{person}{Dominic Abrams}.} \bibinfo{year}{2007}\natexlab{}.
\newblock \showarticletitle{Social cognition and attitudes}.
\newblock In \bibinfo{booktitle}{\emph{Psychology. Third Edition}},
  \bibfield{editor}{\bibinfo{person}{G.~Neil Martin}, \bibinfo{person}{Neil~R.
  Carlson}, {and} \bibinfo{person}{William Buskist}} (Eds.).
  \bibinfo{publisher}{Pearson Education Limited}, \bibinfo{pages}{684--721}.
\newblock
\urldef\tempurl%
\url{https://kar.kent.ac.uk/23659/}
\showURL{%
\tempurl}


\bibitem[\protect\citeauthoryear{Jenkins, Russell, and Suci}{Jenkins
  et~al\mbox{.}}{1958}]%
        {jenkins1958atlas}
\bibfield{author}{\bibinfo{person}{James~J Jenkins}, \bibinfo{person}{Wallace~A
  Russell}, {and} \bibinfo{person}{George~J Suci}.}
  \bibinfo{year}{1958}\natexlab{}.
\newblock \showarticletitle{An atlas of semantic profiles for 360 words}.
\newblock \bibinfo{journal}{\emph{The American Journal of Psychology}}
  \bibinfo{volume}{71}, \bibinfo{number}{4} (\bibinfo{year}{1958}),
  \bibinfo{pages}{688--699}.
\newblock


\bibitem[\protect\citeauthoryear{Jolliffe}{Jolliffe}{1986}]%
        {alma9924548563604107}
\bibfield{author}{\bibinfo{person}{I.~T. Jolliffe}.}
  \bibinfo{year}{1986}\natexlab{}.
\newblock \bibinfo{booktitle}{\emph{Principal component analysis}}.
\newblock \bibinfo{publisher}{Springer-Verlag}, \bibinfo{address}{New York}.
\newblock
\showISBNx{0387962697}
\showLCCN{85027882}


\bibitem[\protect\citeauthoryear{Kozlowski, Taddy, and Evans}{Kozlowski
  et~al\mbox{.}}{2019}]%
        {doi:10.1177/0003122419877135}
\bibfield{author}{\bibinfo{person}{Austin~C. Kozlowski}, \bibinfo{person}{Matt
  Taddy}, {and} \bibinfo{person}{James~A. Evans}.}
  \bibinfo{year}{2019}\natexlab{}.
\newblock \showarticletitle{The Geometry of Culture: Analyzing the Meanings of
  Class through Word Embeddings}.
\newblock \bibinfo{journal}{\emph{American Sociological Review}}
  \bibinfo{volume}{84}, \bibinfo{number}{5} (\bibinfo{year}{2019}),
  \bibinfo{pages}{905--949}.
\newblock
\urldef\tempurl%
\url{https://doi.org/10.1177/0003122419877135}
\showDOI{\tempurl}
\showeprint{https://doi.org/10.1177/0003122419877135}


\bibitem[\protect\citeauthoryear{Kurita, Vyas, Pareek, Black, and
  Tsvetkov}{Kurita et~al\mbox{.}}{2019}]%
        {kurita2019measuring}
\bibfield{author}{\bibinfo{person}{Keita Kurita}, \bibinfo{person}{Nidhi Vyas},
  \bibinfo{person}{Ayush Pareek}, \bibinfo{person}{Alan~W Black}, {and}
  \bibinfo{person}{Yulia Tsvetkov}.} \bibinfo{year}{2019}\natexlab{}.
\newblock \bibinfo{title}{Measuring Bias in Contextualized Word
  Representations}.
\newblock
\newblock
\showeprint[arxiv]{1906.07337}~[cs.CL]


\bibitem[\protect\citeauthoryear{Lan, Chen, Goodman, Gimpel, Sharma, and
  Soricut}{Lan et~al\mbox{.}}{2019}]%
        {lan2019albert}
\bibfield{author}{\bibinfo{person}{Zhenzhong Lan}, \bibinfo{person}{Mingda
  Chen}, \bibinfo{person}{Sebastian Goodman}, \bibinfo{person}{Kevin Gimpel},
  \bibinfo{person}{Piyush Sharma}, {and} \bibinfo{person}{Radu Soricut}.}
  \bibinfo{year}{2019}\natexlab{}.
\newblock \showarticletitle{Albert: A lite bert for self-supervised learning of
  language representations}.
\newblock \bibinfo{journal}{\emph{arXiv preprint arXiv:1909.11942}}
  (\bibinfo{year}{2019}).
\newblock


\bibitem[\protect\citeauthoryear{Leino, Black, Fredrikson, Sen, and
  Datta}{Leino et~al\mbox{.}}{2019}]%
        {leino2019featurewise}
\bibfield{author}{\bibinfo{person}{Klas Leino}, \bibinfo{person}{Emily Black},
  \bibinfo{person}{Matt Fredrikson}, \bibinfo{person}{Shayak Sen}, {and}
  \bibinfo{person}{Anupam Datta}.} \bibinfo{year}{2019}\natexlab{}.
\newblock \bibinfo{title}{Feature-Wise Bias Amplification}.
\newblock
\newblock
\showeprint[arxiv]{1812.08999}~[cs.LG]


\bibitem[\protect\citeauthoryear{Lewis and Lupyan}{Lewis and Lupyan}{2020}]%
        {LewisMolly2020Gsar}
\bibfield{author}{\bibinfo{person}{Molly Lewis} {and} \bibinfo{person}{Gary
  Lupyan}.} \bibinfo{year}{2020}\natexlab{}.
\newblock \showarticletitle{Gender stereotypes are reflected in the
  distributional structure of 25 languages}.
\newblock \bibinfo{journal}{\emph{Nature human behaviour}} \bibinfo{volume}{4},
  \bibinfo{number}{10} (\bibinfo{year}{2020}), \bibinfo{pages}{1021--1028}.
\newblock
\showISSN{2397-3374}


\bibitem[\protect\citeauthoryear{Liang, Li, Zheng, Lim, Salakhutdinov, and
  Morency}{Liang et~al\mbox{.}}{2020}]%
        {liang-etal-2020-towards}
\bibfield{author}{\bibinfo{person}{Paul~Pu Liang},
  \bibinfo{person}{Irene~Mengze Li}, \bibinfo{person}{Emily Zheng},
  \bibinfo{person}{Yao~Chong Lim}, \bibinfo{person}{Ruslan Salakhutdinov},
  {and} \bibinfo{person}{Louis-Philippe Morency}.}
  \bibinfo{year}{2020}\natexlab{}.
\newblock \showarticletitle{Towards Debiasing Sentence Representations}. In
  \bibinfo{booktitle}{\emph{Proceedings of the 58th Annual Meeting of the
  Association for Computational Linguistics}}. \bibinfo{publisher}{Association
  for Computational Linguistics}, \bibinfo{address}{Online},
  \bibinfo{pages}{5502--5515}.
\newblock
\urldef\tempurl%
\url{https://doi.org/10.18653/v1/2020.acl-main.488}
\showDOI{\tempurl}


\bibitem[\protect\citeauthoryear{Lin, Michel, Lieberman, Orwant, Brockman, and
  Petrov}{Lin et~al\mbox{.}}{2012}]%
        {lin2012syntactic}
\bibfield{author}{\bibinfo{person}{Yuri Lin}, \bibinfo{person}{Jean-Baptiste
  Michel}, \bibinfo{person}{Erez~Aiden Lieberman}, \bibinfo{person}{Jon
  Orwant}, \bibinfo{person}{Will Brockman}, {and} \bibinfo{person}{Slav
  Petrov}.} \bibinfo{year}{2012}\natexlab{}.
\newblock \showarticletitle{Syntactic annotations for the google books ngram
  corpus}. In \bibinfo{booktitle}{\emph{Proceedings of the ACL 2012 system
  demonstrations}}. \bibinfo{pages}{169--174}.
\newblock


\bibitem[\protect\citeauthoryear{Liu, Ott, Goyal, Du, Joshi, Chen, Levy, Lewis,
  Zettlemoyer, and Stoyanov}{Liu et~al\mbox{.}}{2019}]%
        {DBLP:journals/corr/abs-1907-11692}
\bibfield{author}{\bibinfo{person}{Yinhan Liu}, \bibinfo{person}{Myle Ott},
  \bibinfo{person}{Naman Goyal}, \bibinfo{person}{Jingfei Du},
  \bibinfo{person}{Mandar Joshi}, \bibinfo{person}{Danqi Chen},
  \bibinfo{person}{Omer Levy}, \bibinfo{person}{Mike Lewis},
  \bibinfo{person}{Luke Zettlemoyer}, {and} \bibinfo{person}{Veselin
  Stoyanov}.} \bibinfo{year}{2019}\natexlab{}.
\newblock \showarticletitle{RoBERTa: {A} Robustly Optimized {BERT} Pretraining
  Approach}.
\newblock \bibinfo{journal}{\emph{CoRR}}  \bibinfo{volume}{abs/1907.11692}
  (\bibinfo{year}{2019}).
\newblock
\showeprint[arxiv]{1907.11692}
\urldef\tempurl%
\url{http://arxiv.org/abs/1907.11692}
\showURL{%
\tempurl}


\bibitem[\protect\citeauthoryear{Manzini, Yao~Chong, Black, and
  Tsvetkov}{Manzini et~al\mbox{.}}{2019}]%
        {manzini-etal-2019-black}
\bibfield{author}{\bibinfo{person}{Thomas Manzini}, \bibinfo{person}{Lim
  Yao~Chong}, \bibinfo{person}{Alan~W Black}, {and} \bibinfo{person}{Yulia
  Tsvetkov}.} \bibinfo{year}{2019}\natexlab{}.
\newblock \showarticletitle{Black is to Criminal as Caucasian is to Police:
  Detecting and Removing Multiclass Bias in Word Embeddings}. In
  \bibinfo{booktitle}{\emph{Proceedings of the 2019 Conference of the North
  {A}merican Chapter of the Association for Computational Linguistics: Human
  Language Technologies, Volume 1 (Long and Short Papers)}}.
  \bibinfo{publisher}{Association for Computational Linguistics},
  \bibinfo{address}{Minneapolis, Minnesota}, \bibinfo{pages}{615--621}.
\newblock
\urldef\tempurl%
\url{https://doi.org/10.18653/v1/N19-1062}
\showDOI{\tempurl}


\bibitem[\protect\citeauthoryear{May, Wang, Bordia, Bowman, and Rudinger}{May
  et~al\mbox{.}}{2019}]%
        {may-etal-2019-measuring}
\bibfield{author}{\bibinfo{person}{Chandler May}, \bibinfo{person}{Alex Wang},
  \bibinfo{person}{Shikha Bordia}, \bibinfo{person}{Samuel~R. Bowman}, {and}
  \bibinfo{person}{Rachel Rudinger}.} \bibinfo{year}{2019}\natexlab{}.
\newblock \showarticletitle{On Measuring Social Biases in Sentence Encoders}.
  In \bibinfo{booktitle}{\emph{Proceedings of the 2019 Conference of the North
  {A}merican Chapter of the Association for Computational Linguistics: Human
  Language Technologies, Volume 1 (Long and Short Papers)}}.
  \bibinfo{publisher}{Association for Computational Linguistics},
  \bibinfo{address}{Minneapolis, Minnesota}, \bibinfo{pages}{622--628}.
\newblock
\urldef\tempurl%
\url{https://doi.org/10.18653/v1/N19-1063}
\showDOI{\tempurl}


\bibitem[\protect\citeauthoryear{Mehrabian and Russell}{Mehrabian and
  Russell}{1974}]%
        {RefWorks:RefID:228-mehrabian1974approach}
\bibfield{author}{\bibinfo{person}{Albert Mehrabian} {and}
  \bibinfo{person}{James~A. Russell}.} \bibinfo{year}{1974}\natexlab{}.
\newblock \bibinfo{booktitle}{\emph{An approach to environmental psychology}}.
\newblock \bibinfo{publisher}{The MIT Press}, \bibinfo{address}{Cambridge, MA,
  US}. xii, 266 pages.
\newblock
\showISBNx{0262130904}
\newblock
\shownote{ID: 1974-22049-000.}


\bibitem[\protect\citeauthoryear{Mei, Fereidooni, and Caliskan}{Mei
  et~al\mbox{.}}{2023}]%
        {mei2023bias}
\bibfield{author}{\bibinfo{person}{Katelyn Mei}, \bibinfo{person}{Sonia
  Fereidooni}, {and} \bibinfo{person}{Aylin Caliskan}.}
  \bibinfo{year}{2023}\natexlab{}.
\newblock \showarticletitle{Bias Against 93 Stigmatized Groups in Masked
  Language Models and Downstream Sentiment Classification Tasks}. In
  \bibinfo{booktitle}{\emph{Proceedings of the 2023 ACM Conference on Fairness,
  Accountability, and Transparency}}. \bibinfo{pages}{1699--1710}.
\newblock


\bibitem[\protect\citeauthoryear{Mikolov, Grave, Bojanowski, Puhrsch, and
  Joulin}{Mikolov et~al\mbox{.}}{2018}]%
        {mikolov2018advances}
\bibfield{author}{\bibinfo{person}{Tomas Mikolov}, \bibinfo{person}{Edouard
  Grave}, \bibinfo{person}{Piotr Bojanowski}, \bibinfo{person}{Christian
  Puhrsch}, {and} \bibinfo{person}{Armand Joulin}.}
  \bibinfo{year}{2018}\natexlab{}.
\newblock \showarticletitle{Advances in Pre-Training Distributed Word
  Representations}. In \bibinfo{booktitle}{\emph{Proceedings of the
  International Conference on Language Resources and Evaluation (LREC 2018)}}.
\newblock


\bibitem[\protect\citeauthoryear{Mikolov, Sutskever, Chen, Corrado, and
  Dean}{Mikolov et~al\mbox{.}}{2013a}]%
        {mikolov2013distributed}
\bibfield{author}{\bibinfo{person}{Tomas Mikolov}, \bibinfo{person}{Ilya
  Sutskever}, \bibinfo{person}{Kai Chen}, \bibinfo{person}{Greg Corrado}, {and}
  \bibinfo{person}{Jeffrey Dean}.} \bibinfo{year}{2013}\natexlab{a}.
\newblock \bibinfo{title}{Distributed Representations of Words and Phrases and
  their Compositionality}.
\newblock
\newblock
\showeprint[arxiv]{1310.4546}~[cs.CL]


\bibitem[\protect\citeauthoryear{Mikolov, Yih, and Zweig}{Mikolov
  et~al\mbox{.}}{2013b}]%
        {mikolov-etal-2013-linguistic}
\bibfield{author}{\bibinfo{person}{Tomas Mikolov}, \bibinfo{person}{Wen-tau
  Yih}, {and} \bibinfo{person}{Geoffrey Zweig}.}
  \bibinfo{year}{2013}\natexlab{b}.
\newblock \showarticletitle{Linguistic Regularities in Continuous Space Word
  Representations}. In \bibinfo{booktitle}{\emph{Proceedings of the 2013
  Conference of the North {A}merican Chapter of the Association for
  Computational Linguistics: Human Language Technologies}}.
  \bibinfo{publisher}{Association for Computational Linguistics},
  \bibinfo{address}{Atlanta, Georgia}, \bibinfo{pages}{746--751}.
\newblock
\urldef\tempurl%
\url{https://aclanthology.org/N13-1090}
\showURL{%
\tempurl}


\bibitem[\protect\citeauthoryear{Mohammad}{Mohammad}{2018}]%
        {vad-acl2018}
\bibfield{author}{\bibinfo{person}{Saif~M. Mohammad}.}
  \bibinfo{year}{2018}\natexlab{}.
\newblock \showarticletitle{Obtaining Reliable Human Ratings of Valence,
  Arousal, and Dominance for 20,000 English Words}. In
  \bibinfo{booktitle}{\emph{Proceedings of The Annual Conference of the
  Association for Computational Linguistics (ACL)}}.
  \bibinfo{address}{Melbourne, Australia}.
\newblock


\bibitem[\protect\citeauthoryear{Nadeem, Bethke, and Reddy}{Nadeem
  et~al\mbox{.}}{2020}]%
        {nadeem2020stereoset}
\bibfield{author}{\bibinfo{person}{Moin Nadeem}, \bibinfo{person}{Anna Bethke},
  {and} \bibinfo{person}{Siva Reddy}.} \bibinfo{year}{2020}\natexlab{}.
\newblock \showarticletitle{Stereoset: Measuring stereotypical bias in
  pretrained language models}.
\newblock \bibinfo{journal}{\emph{arXiv preprint arXiv:2004.09456}}
  (\bibinfo{year}{2020}).
\newblock


\bibitem[\protect\citeauthoryear{Omrani~Sabbaghi and Caliskan}{Omrani~Sabbaghi
  and Caliskan}{2022}]%
        {omrani2022measuring}
\bibfield{author}{\bibinfo{person}{Shiva Omrani~Sabbaghi} {and}
  \bibinfo{person}{Aylin Caliskan}.} \bibinfo{year}{2022}\natexlab{}.
\newblock \showarticletitle{Measuring Gender Bias in Word Embeddings of
  Gendered Languages Requires Disentangling Grammatical Gender Signals}. In
  \bibinfo{booktitle}{\emph{Proceedings of the 2022 AAAI/ACM Conference on AI,
  Ethics, and Society}}. \bibinfo{pages}{518--531}.
\newblock


\bibitem[\protect\citeauthoryear{Osgood, Suci, and Tannenbaum}{Osgood
  et~al\mbox{.}}{1957a}]%
        {osgood1957measurement}
\bibfield{author}{\bibinfo{person}{C.E. Osgood}, \bibinfo{person}{G.J. Suci},
  {and} \bibinfo{person}{P.H. Tannenbaum}.} \bibinfo{year}{1957}\natexlab{a}.
\newblock \bibinfo{booktitle}{\emph{The Measurement of Meaning}}.
\newblock \bibinfo{publisher}{University of Illinois Press}.
\newblock
\showISBNx{9780252745393}
\showLCCN{56005684}
\urldef\tempurl%
\url{https://books.google.com/books?id=Qj8GeUrKZdAC}
\showURL{%
\tempurl}


\bibitem[\protect\citeauthoryear{Osgood, Suci, and Tannenbaum}{Osgood
  et~al\mbox{.}}{1957b}]%
        {RefWorks:RefID:230-osgood1957measurement}
\bibfield{author}{\bibinfo{person}{Charles~E. Osgood},
  \bibinfo{person}{George~J. Suci}, {and} \bibinfo{person}{Percy~H.
  Tannenbaum}.} \bibinfo{year}{1957}\natexlab{b}.
\newblock \bibinfo{booktitle}{\emph{The measurement of meaning}}.
\newblock \bibinfo{publisher}{Univer. Illinois Press},
  \bibinfo{address}{Oxford, England}. 342 pages.
\newblock
\newblock
\shownote{ID: 1958-01561-000.}


\bibitem[\protect\citeauthoryear{Parker, Graff, Kong, Chen, and Maeda}{Parker
  et~al\mbox{.}}{2011}]%
        {parker2011english}
\bibfield{author}{\bibinfo{person}{Robert Parker}, \bibinfo{person}{David
  Graff}, \bibinfo{person}{Junbo Kong}, \bibinfo{person}{Ke Chen}, {and}
  \bibinfo{person}{Kazuaki Maeda}.} \bibinfo{year}{2011}\natexlab{}.
\newblock \showarticletitle{English gigaword fifth edition, linguistic data
  consortium}.
\newblock \bibinfo{journal}{\emph{Google Scholar}} (\bibinfo{year}{2011}).
\newblock


\bibitem[\protect\citeauthoryear{Pennington, Socher, and Manning}{Pennington
  et~al\mbox{.}}{2014}]%
        {pennington-etal-2014-glove}
\bibfield{author}{\bibinfo{person}{Jeffrey Pennington},
  \bibinfo{person}{Richard Socher}, {and} \bibinfo{person}{Christopher
  Manning}.} \bibinfo{year}{2014}\natexlab{}.
\newblock \showarticletitle{{G}lo{V}e: Global Vectors for Word Representation}.
  In \bibinfo{booktitle}{\emph{Proceedings of the 2014 Conference on Empirical
  Methods in Natural Language Processing ({EMNLP})}}.
  \bibinfo{publisher}{Association for Computational Linguistics},
  \bibinfo{address}{Doha, Qatar}, \bibinfo{pages}{1532--1543}.
\newblock
\urldef\tempurl%
\url{https://doi.org/10.3115/v1/D14-1162}
\showDOI{\tempurl}


\bibitem[\protect\citeauthoryear{Peters, Neumann, Iyyer, Gardner, Clark, Lee,
  and Zettlemoyer}{Peters et~al\mbox{.}}{2018a}]%
        {peters2018deep}
\bibfield{author}{\bibinfo{person}{Matthew~E. Peters}, \bibinfo{person}{Mark
  Neumann}, \bibinfo{person}{Mohit Iyyer}, \bibinfo{person}{Matt Gardner},
  \bibinfo{person}{Christopher Clark}, \bibinfo{person}{Kenton Lee}, {and}
  \bibinfo{person}{Luke Zettlemoyer}.} \bibinfo{year}{2018}\natexlab{a}.
\newblock \bibinfo{title}{Deep contextualized word representations}.
\newblock
\newblock
\showeprint[arxiv]{1802.05365}~[cs.CL]


\bibitem[\protect\citeauthoryear{Peters, Neumann, Iyyer, Gardner, Clark, Lee,
  and Zettlemoyer}{Peters et~al\mbox{.}}{2018b}]%
        {peters-etal-2018-deep}
\bibfield{author}{\bibinfo{person}{Matthew~E. Peters}, \bibinfo{person}{Mark
  Neumann}, \bibinfo{person}{Mohit Iyyer}, \bibinfo{person}{Matt Gardner},
  \bibinfo{person}{Christopher Clark}, \bibinfo{person}{Kenton Lee}, {and}
  \bibinfo{person}{Luke Zettlemoyer}.} \bibinfo{year}{2018}\natexlab{b}.
\newblock \showarticletitle{Deep Contextualized Word Representations}. In
  \bibinfo{booktitle}{\emph{Proceedings of the 2018 Conference of the North
  {A}merican Chapter of the Association for Computational Linguistics: Human
  Language Technologies, Volume 1 (Long Papers)}}.
  \bibinfo{publisher}{Association for Computational Linguistics},
  \bibinfo{address}{New Orleans, Louisiana}, \bibinfo{pages}{2227--2237}.
\newblock
\urldef\tempurl%
\url{https://doi.org/10.18653/v1/N18-1202}
\showDOI{\tempurl}


\bibitem[\protect\citeauthoryear{Radford and Narasimhan}{Radford and
  Narasimhan}{2018}]%
        {Radford2018ImprovingLU}
\bibfield{author}{\bibinfo{person}{Alec Radford} {and} \bibinfo{person}{Karthik
  Narasimhan}.} \bibinfo{year}{2018}\natexlab{}.
\newblock \showarticletitle{Improving Language Understanding by Generative
  Pre-Training}.
\newblock


\bibitem[\protect\citeauthoryear{Radford, Wu, Child, Luan, Amodei, and
  Sutskever}{Radford et~al\mbox{.}}{2019}]%
        {radford2019language}
\bibfield{author}{\bibinfo{person}{Alec Radford}, \bibinfo{person}{Jeff Wu},
  \bibinfo{person}{Rewon Child}, \bibinfo{person}{David Luan},
  \bibinfo{person}{Dario Amodei}, {and} \bibinfo{person}{Ilya Sutskever}.}
  \bibinfo{year}{2019}\natexlab{}.
\newblock \showarticletitle{Language Models are Unsupervised Multitask
  Learners}.
\newblock  (\bibinfo{year}{2019}).
\newblock


\bibitem[\protect\citeauthoryear{Raffel, Shazeer, Roberts, Lee, Narang, Matena,
  Zhou, Li, and Liu}{Raffel et~al\mbox{.}}{2020}]%
        {JMLR:v21:20-074}
\bibfield{author}{\bibinfo{person}{Colin Raffel}, \bibinfo{person}{Noam
  Shazeer}, \bibinfo{person}{Adam Roberts}, \bibinfo{person}{Katherine Lee},
  \bibinfo{person}{Sharan Narang}, \bibinfo{person}{Michael Matena},
  \bibinfo{person}{Yanqi Zhou}, \bibinfo{person}{Wei Li}, {and}
  \bibinfo{person}{Peter~J. Liu}.} \bibinfo{year}{2020}\natexlab{}.
\newblock \showarticletitle{Exploring the Limits of Transfer Learning with a
  Unified Text-to-Text Transformer}.
\newblock \bibinfo{journal}{\emph{Journal of Machine Learning Research}}
  \bibinfo{volume}{21}, \bibinfo{number}{140} (\bibinfo{year}{2020}),
  \bibinfo{pages}{1--67}.
\newblock
\urldef\tempurl%
\url{http://jmlr.org/papers/v21/20-074.html}
\showURL{%
\tempurl}


\bibitem[\protect\citeauthoryear{Ravfogel, Elazar, Gonen, Twiton, and
  Goldberg}{Ravfogel et~al\mbox{.}}{2020}]%
        {DBLP:conf/acl/RavfogelEGTG20}
\bibfield{author}{\bibinfo{person}{Shauli Ravfogel}, \bibinfo{person}{Yanai
  Elazar}, \bibinfo{person}{Hila Gonen}, \bibinfo{person}{Michael Twiton},
  {and} \bibinfo{person}{Yoav Goldberg}.} \bibinfo{year}{2020}\natexlab{}.
\newblock \showarticletitle{Null It Out: Guarding Protected Attributes by
  Iterative Nullspace Projection}. In \bibinfo{booktitle}{\emph{Proceedings of
  the 58th Annual Meeting of the Association for Computational Linguistics,
  {ACL} 2020, Online, July 5-10, 2020}}, \bibfield{editor}{\bibinfo{person}{Dan
  Jurafsky}, \bibinfo{person}{Joyce Chai}, \bibinfo{person}{Natalie Schluter},
  {and} \bibinfo{person}{Joel~R. Tetreault}} (Eds.).
  \bibinfo{publisher}{Association for Computational Linguistics},
  \bibinfo{pages}{7237--7256}.
\newblock
\urldef\tempurl%
\url{https://www.aclweb.org/anthology/2020.acl-main.647/}
\showURL{%
\tempurl}


\bibitem[\protect\citeauthoryear{Sheng, Chang, Natarajan, and Peng}{Sheng
  et~al\mbox{.}}{2019}]%
        {sheng2019woman}
\bibfield{author}{\bibinfo{person}{Emily Sheng}, \bibinfo{person}{Kai-Wei
  Chang}, \bibinfo{person}{Prem Natarajan}, {and} \bibinfo{person}{Nanyun
  Peng}.} \bibinfo{year}{2019}\natexlab{}.
\newblock \showarticletitle{The Woman Worked as a Babysitter: On Biases in
  Language Generation}. In \bibinfo{booktitle}{\emph{Proceedings of the 2019
  Conference on Empirical Methods in Natural Language Processing and the 9th
  International Joint Conference on Natural Language Processing
  (EMNLP-IJCNLP)}}. \bibinfo{pages}{3407--3412}.
\newblock


\bibitem[\protect\citeauthoryear{Soler and Apidianaki}{Soler and
  Apidianaki}{2021}]%
        {soler2021let}
\bibfield{author}{\bibinfo{person}{Aina~Gari Soler} {and}
  \bibinfo{person}{Marianna Apidianaki}.} \bibinfo{year}{2021}\natexlab{}.
\newblock \showarticletitle{Let's Play Mono-Poly: BERT Can Reveal Words'
  Polysemy Level and Partitionability into Senses}.
\newblock \bibinfo{journal}{\emph{Transactions of the Association for
  Computational Linguistics (TACL)}} (\bibinfo{year}{2021}).
\newblock


\bibitem[\protect\citeauthoryear{Swinger, De-Arteaga, au2, Leiserson, and
  Kalai}{Swinger et~al\mbox{.}}{2019}]%
        {swinger2019biases}
\bibfield{author}{\bibinfo{person}{Nathaniel Swinger}, \bibinfo{person}{Maria
  De-Arteaga}, \bibinfo{person}{Neil Thomas Heffernan~IV au2},
  \bibinfo{person}{Mark~DM Leiserson}, {and} \bibinfo{person}{Adam~Tauman
  Kalai}.} \bibinfo{year}{2019}\natexlab{}.
\newblock \bibinfo{title}{What are the biases in my word embedding?}
\newblock
\newblock
\showeprint[arxiv]{1812.08769}~[cs.CL]


\bibitem[\protect\citeauthoryear{Tan and Celis}{Tan and Celis}{2019}]%
        {DBLP:conf/nips/TanC19}
\bibfield{author}{\bibinfo{person}{Yi~Chern Tan} {and}
  \bibinfo{person}{L.~Elisa Celis}.} \bibinfo{year}{2019}\natexlab{}.
\newblock \showarticletitle{Assessing Social and Intersectional Biases in
  Contextualized Word Representations}. In \bibinfo{booktitle}{\emph{Advances
  in Neural Information Processing Systems 32: Annual Conference on Neural
  Information Processing Systems 2019, NeurIPS 2019, December 8-14, 2019,
  Vancouver, BC, Canada}}, \bibfield{editor}{\bibinfo{person}{Hanna~M.
  Wallach}, \bibinfo{person}{Hugo Larochelle}, \bibinfo{person}{Alina
  Beygelzimer}, \bibinfo{person}{Florence d'Alch{\'{e}}{-}Buc},
  \bibinfo{person}{Emily~B. Fox}, {and} \bibinfo{person}{Roman Garnett}}
  (Eds.). \bibinfo{pages}{13209--13220}.
\newblock
\urldef\tempurl%
\url{https://proceedings.neurips.cc/paper/2019/hash/201d546992726352471cfea6b0df0a48-Abstract.html}
\showURL{%
\tempurl}


\bibitem[\protect\citeauthoryear{Tellegen}{Tellegen}{1985}]%
        {RefWorks:RefID:229-tellegen1985structures}
\bibfield{author}{\bibinfo{person}{Auke Tellegen}.}
  \bibinfo{year}{1985}\natexlab{}.
\newblock \bibinfo{booktitle}{\emph{Structures of mood and personality and
  their relevance to assessing anxiety, with an emphasis on self-report}}.
\newblock \bibinfo{publisher}{Lawrence Erlbaum Associates, Inc},
  \bibinfo{address}{Hillsdale, NJ, US}, \bibinfo{pages}{681--706}.
\newblock
\showISBNx{0-89859-532-0 (Hardcover)}
\newblock
\shownote{ID: 1985-97708-037.}


\bibitem[\protect\citeauthoryear{Timkey and van Schijndel}{Timkey and van
  Schijndel}{2021}]%
        {timkey2021all}
\bibfield{author}{\bibinfo{person}{William Timkey} {and}
  \bibinfo{person}{Marten van Schijndel}.} \bibinfo{year}{2021}\natexlab{}.
\newblock \showarticletitle{All Bark and No Bite: Rogue Dimensions in
  Transformer Language Models Obscure Representational Quality}. In
  \bibinfo{booktitle}{\emph{Proceedings of the 2021 Conference on Empirical
  Methods in Natural Language Processing}}. \bibinfo{pages}{4527--4546}.
\newblock


\bibitem[\protect\citeauthoryear{Toney and Caliskan}{Toney and
  Caliskan}{2020}]%
        {toney2020valnorm}
\bibfield{author}{\bibinfo{person}{Autumn Toney} {and} \bibinfo{person}{Aylin
  Caliskan}.} \bibinfo{year}{2020}\natexlab{}.
\newblock \bibinfo{title}{ValNorm Quantifies Semantics to Reveal Consistent
  Valence Biases Across Languages and Over Centuries}.
\newblock
\newblock
\showeprint[arxiv]{2006.03950}~[cs.CY]


\bibitem[\protect\citeauthoryear{Toney, Pandey, Guo, Broniatowski, and
  Caliskan}{Toney et~al\mbox{.}}{2021}]%
        {toney2021automatically}
\bibfield{author}{\bibinfo{person}{Autumn Toney}, \bibinfo{person}{Akshat
  Pandey}, \bibinfo{person}{Wei Guo}, \bibinfo{person}{David Broniatowski},
  {and} \bibinfo{person}{Aylin Caliskan}.} \bibinfo{year}{2021}\natexlab{}.
\newblock \showarticletitle{Automatically characterizing targeted information
  operations through biases present in discourse on twitter}. In
  \bibinfo{booktitle}{\emph{2021 IEEE 15th International Conference on Semantic
  Computing (ICSC)}}. IEEE, \bibinfo{pages}{82--83}.
\newblock


\bibitem[\protect\citeauthoryear{Trinh and Le}{Trinh and Le}{2018}]%
        {trinh2018simple}
\bibfield{author}{\bibinfo{person}{Trieu~H Trinh} {and} \bibinfo{person}{Quoc~V
  Le}.} \bibinfo{year}{2018}\natexlab{}.
\newblock \showarticletitle{A simple method for commonsense reasoning}.
\newblock \bibinfo{journal}{\emph{arXiv preprint arXiv:1806.02847}}
  (\bibinfo{year}{2018}).
\newblock


\bibitem[\protect\citeauthoryear{Vaswani, Shazeer, Parmar, Uszkoreit, Jones,
  Gomez, Kaiser, and Polosukhin}{Vaswani et~al\mbox{.}}{2017}]%
        {vaswani2017attention}
\bibfield{author}{\bibinfo{person}{Ashish Vaswani}, \bibinfo{person}{Noam
  Shazeer}, \bibinfo{person}{Niki Parmar}, \bibinfo{person}{Jakob Uszkoreit},
  \bibinfo{person}{Llion Jones}, \bibinfo{person}{Aidan~N Gomez},
  \bibinfo{person}{{\L}ukasz Kaiser}, {and} \bibinfo{person}{Illia
  Polosukhin}.} \bibinfo{year}{2017}\natexlab{}.
\newblock \showarticletitle{Attention is all you need}. In
  \bibinfo{booktitle}{\emph{Advances in neural information processing
  systems}}. \bibinfo{pages}{5998--6008}.
\newblock


\bibitem[\protect\citeauthoryear{Warriner, Kuperman, and Brysbaert}{Warriner
  et~al\mbox{.}}{2013}]%
        {warriner2013norms}
\bibfield{author}{\bibinfo{person}{Amy~Beth Warriner}, \bibinfo{person}{Victor
  Kuperman}, {and} \bibinfo{person}{Marc Brysbaert}.}
  \bibinfo{year}{2013}\natexlab{}.
\newblock \showarticletitle{Norms of valence, arousal, and dominance for 13,915
  English lemmas}.
\newblock \bibinfo{journal}{\emph{Behavior research methods}}
  \bibinfo{volume}{45}, \bibinfo{number}{4} (\bibinfo{year}{2013}),
  \bibinfo{pages}{1191--1207}.
\newblock


\bibitem[\protect\citeauthoryear{Wolf, Chaumond, Debut, Sanh, Delangue, Moi,
  Cistac, Funtowicz, Davison, Shleifer, et~al\mbox{.}}{Wolf
  et~al\mbox{.}}{2020}]%
        {wolf2020transformers}
\bibfield{author}{\bibinfo{person}{Thomas Wolf}, \bibinfo{person}{Julien
  Chaumond}, \bibinfo{person}{Lysandre Debut}, \bibinfo{person}{Victor Sanh},
  \bibinfo{person}{Clement Delangue}, \bibinfo{person}{Anthony Moi},
  \bibinfo{person}{Pierric Cistac}, \bibinfo{person}{Morgan Funtowicz},
  \bibinfo{person}{Joe Davison}, \bibinfo{person}{Sam Shleifer},
  {et~al\mbox{.}}} \bibinfo{year}{2020}\natexlab{}.
\newblock \showarticletitle{Transformers: State-of-the-art natural language
  processing}. In \bibinfo{booktitle}{\emph{Proceedings of the 2020 Conference
  on Empirical Methods in Natural Language Processing: System Demonstrations}}.
  \bibinfo{pages}{38--45}.
\newblock


\bibitem[\protect\citeauthoryear{Wolfe and Caliskan}{Wolfe and
  Caliskan}{2021}]%
        {wolfe2021low}
\bibfield{author}{\bibinfo{person}{Robert Wolfe} {and} \bibinfo{person}{Aylin
  Caliskan}.} \bibinfo{year}{2021}\natexlab{}.
\newblock \bibinfo{title}{Low Frequency Names Exhibit Bias and Overfitting in
  Contextualizing Language Models}.
\newblock
\newblock
\showeprint[arxiv]{2110.00672}~[cs.CY]


\bibitem[\protect\citeauthoryear{Wolfe and Caliskan}{Wolfe and
  Caliskan}{2022a}]%
        {wolfe2022american}
\bibfield{author}{\bibinfo{person}{Robert Wolfe} {and} \bibinfo{person}{Aylin
  Caliskan}.} \bibinfo{year}{2022}\natexlab{a}.
\newblock \showarticletitle{American== white in multimodal language-and-image
  ai}. In \bibinfo{booktitle}{\emph{Proceedings of the 2022 AAAI/ACM Conference
  on AI, Ethics, and Society}}. \bibinfo{pages}{800--812}.
\newblock


\bibitem[\protect\citeauthoryear{Wolfe and Caliskan}{Wolfe and
  Caliskan}{2022b}]%
        {wolfe2022detecting}
\bibfield{author}{\bibinfo{person}{Robert Wolfe} {and} \bibinfo{person}{Aylin
  Caliskan}.} \bibinfo{year}{2022}\natexlab{b}.
\newblock \showarticletitle{Detecting Emerging Associations and Behaviors With
  Regional and Diachronic Word Embeddings}. In \bibinfo{booktitle}{\emph{2022
  IEEE 16th International Conference on Semantic Computing (ICSC)}}. IEEE,
  \bibinfo{pages}{91--98}.
\newblock


\bibitem[\protect\citeauthoryear{Wolfe and Caliskan}{Wolfe and
  Caliskan}{2022c}]%
        {wolfe2022markedness}
\bibfield{author}{\bibinfo{person}{Robert Wolfe} {and} \bibinfo{person}{Aylin
  Caliskan}.} \bibinfo{year}{2022}\natexlab{c}.
\newblock \showarticletitle{Markedness in visual semantic ai}. In
  \bibinfo{booktitle}{\emph{Proceedings of the 2022 ACM Conference on Fairness,
  Accountability, and Transparency}}. \bibinfo{pages}{1269--1279}.
\newblock


\bibitem[\protect\citeauthoryear{Wolfe and Caliskan}{Wolfe and
  Caliskan}{2022d}]%
        {wolfe2022vast}
\bibfield{author}{\bibinfo{person}{Robert Wolfe} {and} \bibinfo{person}{Aylin
  Caliskan}.} \bibinfo{year}{2022}\natexlab{d}.
\newblock \showarticletitle{{V}{A}{S}{T}: The Valence-Assessing Semantics Test
  for Contextualizing Language Models}.
\newblock \bibinfo{journal}{\emph{Association for the Advancement of Artificial
  Intelligence (AAAI)}} (\bibinfo{year}{2022}).
\newblock


\bibitem[\protect\citeauthoryear{Yang, Dai, Yang, Carbonell, Salakhutdinov, and
  Le}{Yang et~al\mbox{.}}{2019}]%
        {NEURIPS2019_dc6a7e65}
\bibfield{author}{\bibinfo{person}{Zhilin Yang}, \bibinfo{person}{Zihang Dai},
  \bibinfo{person}{Yiming Yang}, \bibinfo{person}{Jaime Carbonell},
  \bibinfo{person}{Russ~R Salakhutdinov}, {and} \bibinfo{person}{Quoc~V Le}.}
  \bibinfo{year}{2019}\natexlab{}.
\newblock \showarticletitle{XLNet: Generalized Autoregressive Pretraining for
  Language Understanding}. In \bibinfo{booktitle}{\emph{Advances in Neural
  Information Processing Systems}},
  \bibfield{editor}{\bibinfo{person}{H.~Wallach},
  \bibinfo{person}{H.~Larochelle}, \bibinfo{person}{A.~Beygelzimer},
  \bibinfo{person}{F.~d\textquotesingle Alch\'{e}-Buc},
  \bibinfo{person}{E.~Fox}, {and} \bibinfo{person}{R.~Garnett}} (Eds.),
  Vol.~\bibinfo{volume}{32}. \bibinfo{publisher}{Curran Associates, Inc.}
\newblock
\urldef\tempurl%
\url{https://proceedings.neurips.cc/paper/2019/file/dc6a7e655d7e5840e66733e9ee67cc69-Paper.pdf}
\showURL{%
\tempurl}


\bibitem[\protect\citeauthoryear{Zhao, Wang, Yatskar, Cotterell, Ordonez, and
  Chang}{Zhao et~al\mbox{.}}{2019}]%
        {zhao-etal-2019-gender}
\bibfield{author}{\bibinfo{person}{Jieyu Zhao}, \bibinfo{person}{Tianlu Wang},
  \bibinfo{person}{Mark Yatskar}, \bibinfo{person}{Ryan Cotterell},
  \bibinfo{person}{Vicente Ordonez}, {and} \bibinfo{person}{Kai-Wei Chang}.}
  \bibinfo{year}{2019}\natexlab{}.
\newblock \showarticletitle{Gender Bias in Contextualized Word Embeddings}. In
  \bibinfo{booktitle}{\emph{Proceedings of the 2019 Conference of the North
  {A}merican Chapter of the Association for Computational Linguistics: Human
  Language Technologies, Volume 1 (Long and Short Papers)}}.
  \bibinfo{publisher}{Association for Computational Linguistics},
  \bibinfo{address}{Minneapolis, Minnesota}, \bibinfo{pages}{629--634}.
\newblock
\urldef\tempurl%
\url{https://doi.org/10.18653/v1/N19-1064}
\showDOI{\tempurl}


\bibitem[\protect\citeauthoryear{Zhao, Wang, Yatskar, Ordonez, and Chang}{Zhao
  et~al\mbox{.}}{2017}]%
        {zhao-etal-2017-men}
\bibfield{author}{\bibinfo{person}{Jieyu Zhao}, \bibinfo{person}{Tianlu Wang},
  \bibinfo{person}{Mark Yatskar}, \bibinfo{person}{Vicente Ordonez}, {and}
  \bibinfo{person}{Kai-Wei Chang}.} \bibinfo{year}{2017}\natexlab{}.
\newblock \showarticletitle{Men Also Like Shopping: Reducing Gender Bias
  Amplification using Corpus-level Constraints}. In
  \bibinfo{booktitle}{\emph{Proceedings of the 2017 Conference on Empirical
  Methods in Natural Language Processing}}. \bibinfo{publisher}{Association for
  Computational Linguistics}, \bibinfo{address}{Copenhagen, Denmark},
  \bibinfo{pages}{2979--2989}.
\newblock
\urldef\tempurl%
\url{https://doi.org/10.18653/v1/D17-1323}
\showDOI{\tempurl}


\bibitem[\protect\citeauthoryear{Zhu, Kiros, Zemel, Salakhutdinov, Urtasun,
  Torralba, and Fidler}{Zhu et~al\mbox{.}}{2015}]%
        {zhu2015aligning}
\bibfield{author}{\bibinfo{person}{Yukun Zhu}, \bibinfo{person}{Ryan Kiros},
  \bibinfo{person}{Rich Zemel}, \bibinfo{person}{Ruslan Salakhutdinov},
  \bibinfo{person}{Raquel Urtasun}, \bibinfo{person}{Antonio Torralba}, {and}
  \bibinfo{person}{Sanja Fidler}.} \bibinfo{year}{2015}\natexlab{}.
\newblock \showarticletitle{Aligning books and movies: Towards story-like
  visual explanations by watching movies and reading books}. In
  \bibinfo{booktitle}{\emph{Proceedings of the IEEE international conference on
  computer vision}}. \bibinfo{pages}{19--27}.
\newblock


\end{thebibliography}

\end{document}